\documentclass{tMPH2eRem}

\usepackage{fixltx2e} 

\begin{document}

\title{Probabilistic Model to Treat Flexibility in Molecular Contacts}

\author{Riku Hakulinen$^{\ast}$\thanks{$^\ast$Corresponding author. Email: riku.hakulinen@gmail.com
\vspace{6pt}} and Santeri Puranen$^{a,b,c}$\\\vspace{6pt}
$^{a}${\em{Department of Computer Science, Aalto University, Espoo, Finland}}
$^{b}${\em{Department of Computer Science, University of Helsinki, Helsinki, Finland}}
$^{c}${\em{Helsinki Institute for Information Technology HIIT, Helsinki, Finland}}
}
\maketitle
\vspace{8pt}

\begin{abstract}

Evaluating accessible conformational space is computationally expensive and thermal 
motions are partly neglected in computer models of molecular interactions. This 
produces error into the estimates of binding strength. We introduce a method for 
modelling interactions so that structural flexibility is inherently taken into account. 
It has a statistical model for 3D properties of \textit{nonlocal} contacts and a physics 
based description of \textit{local} interactions, based on mechanical torque. The form of 
the torque barrier is derived using a representation of the local electronic structure, 
which is presumed to improve transferability, compared to traditional force fields. The 
nonlocal contacts are more distant than 1-4 interactions and Target-atoms are represented 
by 3D probability densities. Probability mass quantifies strength of contact and is 
calculated as an overlap integral. Repulsion is described by negative probability density, 
allowing probability mass to be used as the descriptor of contact preference. As a result, 
we are able to transform the high-dimensional problem into a simpler evaluation of 
three-dimensional integrals. We outline how this scoring function gives a tool to study 
the enthalpy--entropy compensation and demonstrate the feasibility of our approach by 
evaluating numerical probability masses for chosen side chain to main chain contacts 
in a lysine dipeptide structure.

\bigskip

\end{abstract}

\section{Introduction}

A computationally affordable approach that would allow for chemically
accurate simulation of interactions in large protein complexes in a given 
molecular environment, would be decisively useful for the study of biological 
systems. Namely, this would open up new avenues for theoretical analysis of 
biochemical processes and aid in the design of complex molecular components in 
bioscience research. Extensive molecular modeling at the molecular systems level 
is an emerging, multifaceted field and a manifestation of chemical physics 
applied to biology \cite{Gruebele2013}. Molecular simulations are based on 
computational methods that describe molecular interactions and so regulate 
the virtual model of the studied system. There exist several methods that can 
be used to routinely calculate strengths of static interactions for any given 
complex of molecular structures, see e.g. refs\cite{Mura2014,Gao2014}, but 
incorporating such factors as thermal motion and flexibility, that 
are required for the model to be considered realistic, has proven a significant 
challenge. These factors are of great importance in understanding, for example, 
the process of molecular complex formation \cite{Forrey2012} and for predicting 
relative protein conformations \cite{Hernandez2012}. One specific example where the 
detailed understanding of the variability of also the local protein structure has an 
essential role, is given by the function of an ion channel \cite{Allen2004}. We are 
developing a method to treat flexibility and thermal motion in macromolecular systems, 
including protein-ligand interactions. In this work we outline the basic principles of 
the approach and apply the method to a small structure (a lysine dipeptide), accessible 
at the present stage of implementability, see Discussion for details. The current 
difficulties of force fields, solving of which also our method is targeted for, are 
discussed in a somewhat different setting of nanostructures in, e.g., section 2.4 
of ref \cite{Heinz2013}.\\

The method presented here treats internal (local) molecular conformations in terms 
of classical mechanics, but for external (nonlocal) contacts incorporates a concept 
used in quantum theory formalism, namely the overlap integral for functions of 
position, see e.g. \cite[pp.~154-156, 325-326]{Gasiorowicz1996}. Overlap probability 
mass quantifies here the strength of an interaction, and is defined directly based on 
3D probability densities, instead of wave functions which do not appear in this approach. 
We therefore try to approximate the information that is assumed, for example, a quantum 
chemistry  description would produce. At present, this is done through experimental 
coordinate data and prior chemical knowledge. Quantum chemistry results are used as 
reference, though not necessarily directly. Namely, questions concerning the role of 
interactions involving varying electron densities, like dispersion \cite{Martin2013}, are 
at least in the present model considered further than 1-4 interactions and therefore 
implicit in the molecular fragment classification. In general, the fragment classification 
is central to how quantitative this method is, or can be.\\

Strain determines internal preferences in the molecular structure and is in our approach described 
with the classical mechanical moment of force $\bar{r}\times\bar{F}$, i.e., the cross product of a 
position vector with a force vector. It has the unit newton meters (Nm), and quantifies here how 
strongly an internally rotating structure is influenced by the charge distributions present at 
each end of a rotatable bond, so that the system is forced to move towards an equilibrium conformation.
This approach was chosen to treat the flexible molecule as a mechanical system composed 
of levers and pivot points, not masses in space experiencing potentials. Namely, torque 
is considered as a natural quantity for describing a covalent structure. Rotations about 
single bonds are the primary form of motion realizing the structural flexibility considered 
in this work. Rotational barriers over full rotations around rotatable bonds are calculated
based on the moment of force. Adjustable average bond angles and lengths are used, though bending and 
stretching of bonds could be taken into account through the same scheme, by making the 
parameter bond angles and lengths depend on the angle of rotation. Nevertheless, at 
least with respect to the relevant case here, a substituted hydrocarbon straight chain 
segment, an average constant bond angle seems reasonable, because of bond angle stabilizing 
electrostatic interactions over adjacent rotatable bonds. The main goal of this paper 
is to describe a novel approach for modeling molecular structure and interactions, and 
to demonstrate how this probabilistic method is used to obtain chemically relevant and 
commensurate numerical information.\\

The partitioning of model components differs here from a typical force
field \cite{Lindorff-Larsen2010,Mura2014,Mackerell2004,Heinz2013}, in that the energy
landscape of internal rotations for a molecule is analytically further defined.
In practical terms, the dihedral part of a molecular mechanics force field has a partially
predefined functional form, with parameters whose values are derived from
fitting either to experimental data or to quantum chemistry calculations. In 
contrast, the form of the internal torque is here derived using a representation 
of electronic structure, i.e., elementary properties of the electronic structure is 
the source of parameters. This approach is expected to improve the transferability 
\cite{Eggimann2014,Gao2014} of the energy function, as compared to traditional force 
fields. Another important feature is that the structural flexibility 
\cite{Forrey2012,Tzanov2014,Andrusier2008} allowed by the degrees of freedom and 
utilized by thermal energy, is captured in one theoretical object, a three dimensional 
probability density. Using these densities together with the overlap integral method, makes 
it possible to simultaneously describe noncovalent interaction strength and restrictions 
to the freedom of motion.\\

The Method section describes advances made in this study to an existing model 
framework from our previous work \cite{Hakulinen2012,Hakulinen2013}. These in
turn recast the founding work by Rantanen \textit{et al.} \cite{Rantanen2003}.
In section Results, we show how the method has been refined, especially its
ability to capture the fundamentally important molecular flexibility and 
the way this is incorporated in the contact preference calculations. We also
describe how repulsion can be treated as negative probability and outline
enthalpy--entropy compensation as a result of the spatial properties of
interactions. Then, we present numerical results from applying the method to
a test structure. Finally, in the Discussion, we consider potential ways of 
further improving the model.

\section{The Method}
This is a molecular fragment based method. It means here three-atom fragments (each 
belonging to a class in a chemical classification, see e.g. \cite{Hakulinen2012}) that are 
selected from the studied molecules. A reference frame is attached to each chosen fragment, in 
order to model its noncovalent contacts in the system. The reference frame, together 
with pre-determined 3D probability densitities, called contact preference densities, allow
for both distance and direction dependent analysis of the interactions with a chosen molecular 
environment. Final step in estimating the strength of a contact is evaluating an overlap integral. 
The integrand is derived from the contact preference density and a Target-atom distribution, where 
the latter has also been modelled as a 3D probability density. In determining the distribution 
of the Target-atoms, internal preferences of the interacting molecular structures and a thermal 
energy level (or a distribution of levels) are required. The internal preferences consist of 
the torque based strain (local) together with possible intramolecular contacts (nonlocal).

\begin{figure}[ht]
\includegraphics{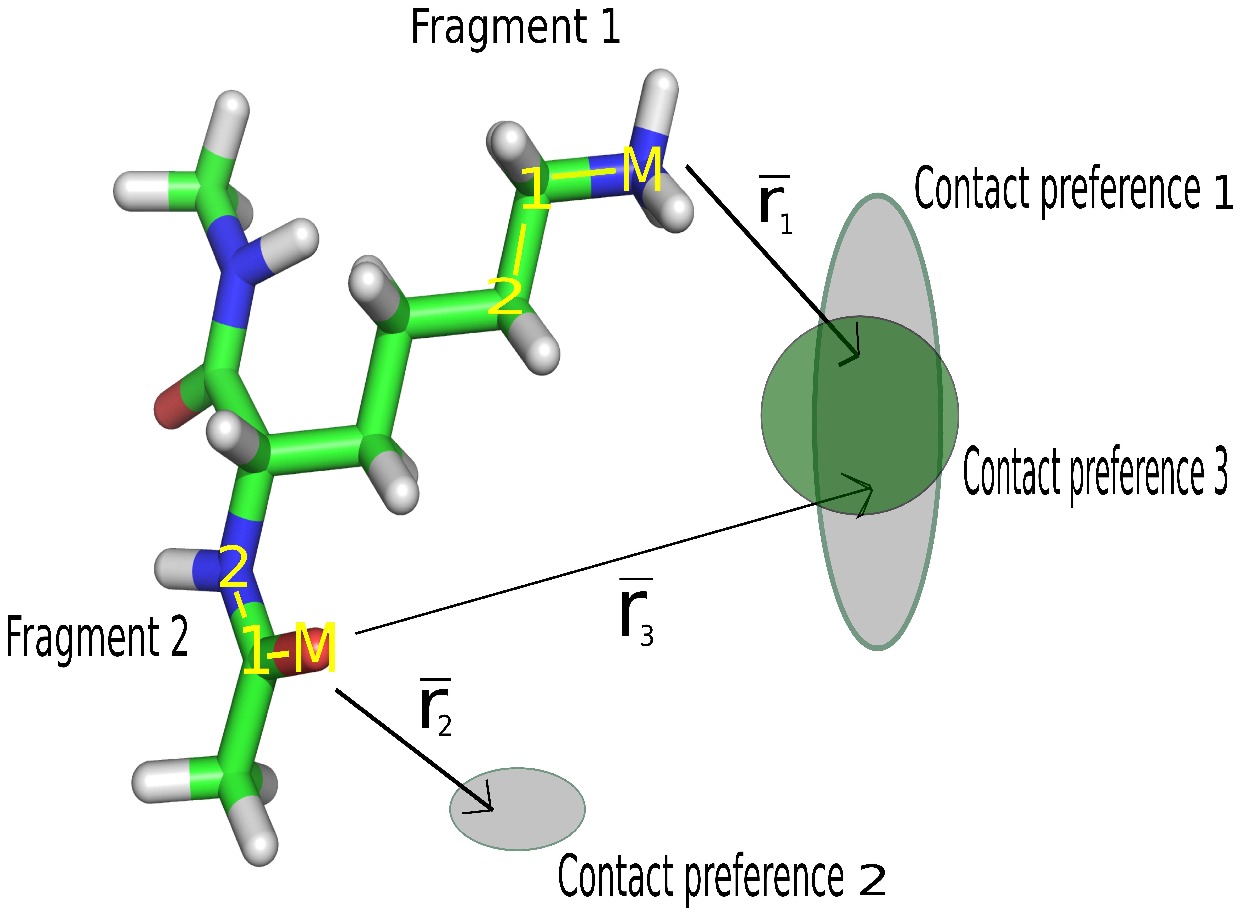}
\caption{Molecular fragments and their contact preferences. In figure shown the structure
 used as a test structure (lysine dipeptide, see text for details). Calculations require three-atom 
fragments (M-1-2, yellow), for which 3D probabilistic contact preferences have been determined. The 
latter shown schematically (filled ellipses) for two fragments, with position vectors from fragment
to a point inside preference density with corresponding indeces (1,2 or 3).}
\label{fig1}
\end{figure} 

\subsection{Reference frame and molecular fragment} Orienting a molecular
fragment in three dimensional space requires three atoms, called here M, 1 and 2. 
This is depicted in Figure \ref{fig1} using the end group of lysine side chain and a 
carbonyl group from the main chain as examples. The former belongs to the fragment 
class for Primary amine nitrogens bonded to an aliphatic structure (class \textbf{f26} 
in \cite{Hakulinen2012}) and the latter to the class for Amide group oxygens bonded to a 
non-aromatic structure (class \textbf{f26} in \cite{Hakulinen2012}). In Figure\ \ref{fig1}, 
the Main-atoms M are NZ and O, using Protein Data Bank atom names, and the two other atoms 
(1 and 2) are, correspondingly, the next two carbons in the side-chain and the other two atoms 
of the amide group (C and N ), as shown. The bonds in the lysine side chain being rotatable, NZ 
can obtain positions from a complicated spatial distribution. A convention used in 
this work for calculations, is that a fragments rotatable bond is always the bond between 
Main-atom M and atom 1, the fragment realization in the studied structure is changed accordingly.\\

A representation of the local electronic structure is required for describing the 
internal local strain in a molecular structure. The representation used in this work 
consists of point charges on atoms and bonds, i.e., in addition to a standard partial
charges scheme, also bonds between atoms are assigned separate charges. This approach is used 
because molecular strain is produced in this model by torque, which  depends on distance 
(along a covalent bond) from the pivot point (an atom). Also, the distance between atoms that are
directly bonded to two adjacent atoms, like the ends of a rotatable bond, is not large compared 
to dimensions of a bond-atom charge distribution, so that the validity of a multipole expansion 
is not obvious for the 1-4 interaction.\\

\begin{figure}[ht]
\includegraphics{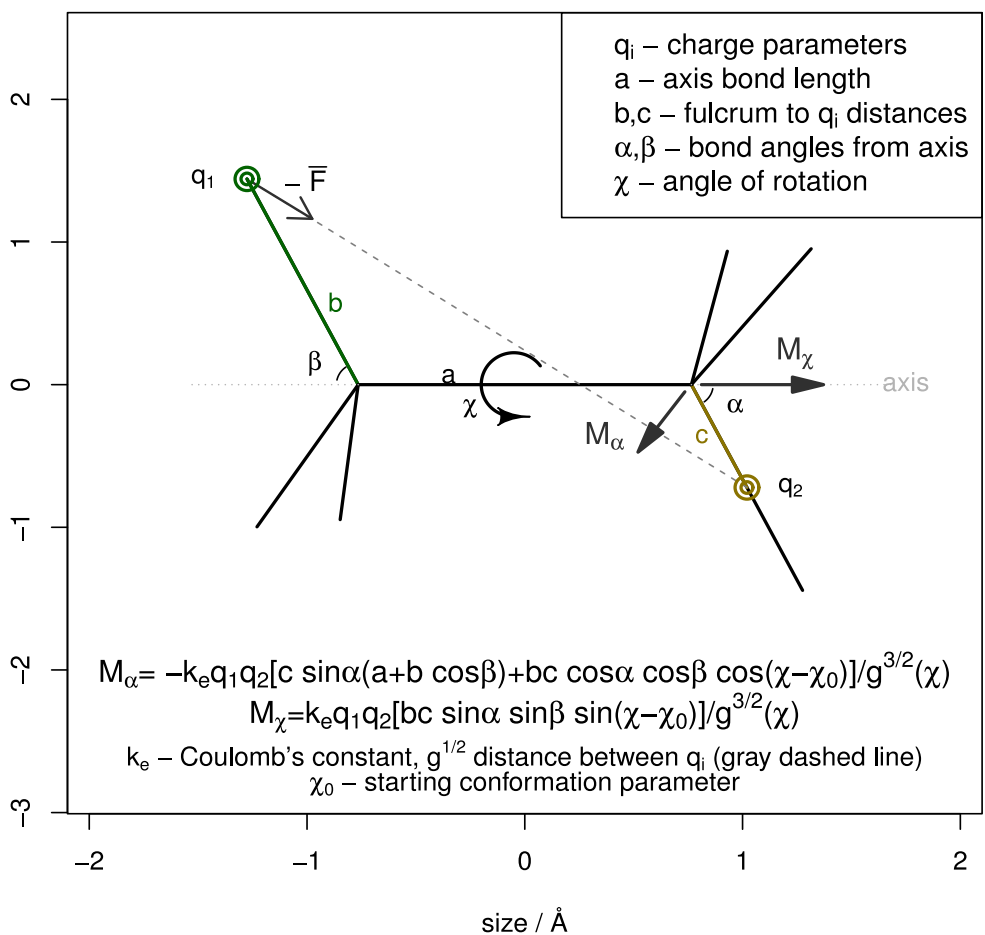}
\caption{Torque parameters for electrostatic interaction over a rotatable
bond. Potential energy is replaced in this method by two components of torque, 
to an angle of rotation unit vector direction ($M_{\alpha}$) and to axis of rotation 
direction ($M_{\chi}$). The total amount of torque causing strain in the structure is a sum
of individual terms shown in the lower part of the graphics. These terms
correspond to parameter charge pairs in covalent bonds ($q_2$) and on atoms
($q_1$), including atom--atom, atom--bond and bond--bond interactions. The depicted
direction of force vector $-\bar{F}$ corresponds to opposite
$q_1$ and $q_2$ charges. $M_{\alpha}$ has one of infinitely many directions
perpendicular to the axis of rotation.}
\label{fig2}
\end{figure}

The immediate local structure, centered around a rotatable bond, is schematically illustrated in 
Figure \ref{fig2}. There the Main-atom (M) and atom 1 would be connected by the rotatable bond $a$, 
and atom 2 could be the one having the parameter charge $q_1$, i.e., one bond away from the axis 
of rotation, on the $\beta$ angle side of the axis bond.

\subsection{Torque model of internal strain}

The rotation barrier structure is in the following determined for a single rotatable 
bond contribution to the background \textit{internal} energy state (strain without 
\textit{nonlocal} contacts). It is calculated as torque due to force couples of 
oscillating classical electrostatic forces produced by parameter charges. As stated, 
these charges are in bonds and on atoms in the vicinity of both ends of the rotatable bond, 
but not as part of it, see Figure\ \ref{fig2}. The interaction can be described using 
Coulomb's inverse square law between the individual charges \cite[pp.~27-37]{Reitz1992}, which 
gives a realistic force estimate (in vacuum), provided that the point charge distribution 
is representative of the interacting spatial charge distributions.\\

In addition to the mechanical suitableness and alleged better transferability, shortly discussed 
in Introduction, another useful aspect is that because the existence of a well defined 
potential for the hindered rotation is not presumed, path-dependence of internal rotations 
could in principle be studied. Namely, forces depending on variables disturbing the definition
of a potential, like velocity ($\bar{F}(\bar{r},\dot{\bar{r}})$), can be used to model suspected 
internal (to structure) friction and dispersion in the \emph{local} component of the method.  

\paragraph{Torque equations} In order to model the energy states during
internal rotation, the combined torque produced by two force components at each
end of the rotatable bond is used. One component is towards the direction of bond 
bending and the other towards the direction of rotation. The corresponding moments 
of force, $M_{\alpha}$ and $M_{\chi}$, are then defined with respect to a pivot 
point (an atom at one end of a rotatable bond) and the axis of rotation (the 
z-axis as defined in this work), see schematic Figure\ \ref{fig2} for details. 
The primary factor expressing variation in internal angular preference is the 
change in the sum of the net values of these torque components, separately added 
over all parameter charge interactions. Using bond and angle naming convention shown 
in Figure\ \ref{fig2}, equation for the (axial) vector torque components is
\begin{equation}
\bar{M}=[\bar{c}\times\bar{F},\bar{b}\times(-\bar{F})]
\label{Meq}
\end{equation}
where a half of the force couple is 
\begin{equation}
\bar{F}=\bar{F}_{\rho}+\bar{F}_{\alpha}+\bar{F}_{\chi}=F_{\rho}\bar{u}_{\rho}+F_{\alpha}\bar{u}_{\alpha}+F_{\chi}\bar{u}_{\chi}=-F^{(b)}_{\rho}\bar{u}^{(b)}_{\rho}-F^{(b)}_{\alpha}\bar{u}^{(b)}_{\alpha}-F^{(b)}_{\chi}\bar{u}^{(b)}_{\chi}
\end{equation}
and $\bar{u}_{\rho}$ is along a rotating bond, $b$ and $c$ in Figure\ \ref{fig2}.
Vector $\bar{u}_{\chi}$ is the unit tangent vector of the
$\chi$-related rotation arch tracked by a parameter charge, $q_i$ in Figure\
\ref{fig2}, and $\bar{u}_{\alpha}$ is the same for bond bending, parameterized by
angles $\alpha$ and $\beta$. The superscript $b$ refers to the components at that
end of the rotatable bond with parameters $b$ and $\beta$. The force components
have the form given in Table \ref{force}, where the multiplier $k(\chi)$, that
contains the angle of rotation dependent squared distance (in square brackets),
has the form
\begin{align}
k_i(\chi)=&k_eq_1q_2[a_{i}^{2}+b_{i}^{2}+c_{i}^{2}+2a_{i}c_{i}\cos(\alpha_{i})+2a_{i}b_{i}\cos(\beta_{i})+2b_{i}c_{i}\cos(\alpha_{i})\cos(\beta_{i})-
\nonumber\\
&2b_{i}c_{i}\sin(\alpha_{i})\sin(\beta_{i})\cos(\chi-\chi^{(i)}_{0})]^{-\frac{3}{2}}\label{distSq}
\end{align}
The definitions given in Figure\ \ref{fig2} for the structural parameters
(${a,b,c,\alpha,\beta,\chi_{0}}$) are in respective order: axis bond length, two 
distances from pivot point, two bond angles and an orientation of one of the bonds 
($\chi_0$ is not rotated, but kept fixed during rotatable bond specific computations), 
and $k_e$ is Coulomb's constant.\\
\begin{table}
\tbl{Force components needed for calculating torque caused by rotation about a bond.}
{\begin{tabular}{| l | l | l |} \hline
 Component & Definition  & Functional form \\
\hline
  $F_{\chi}$ & $\bar{F}\cdot\bar{u}_{\chi}$ &  $k(\chi)b\sin\beta\sin(\chi-\chi_0)$\\
  $F^{(b)}_{\chi}$ & $\bar{F}\cdot\bar{u}^{(b)}_{\chi}$ & $k(\chi)c\sin\alpha\sin(\chi-\chi_0)$\\
  $F_{\alpha}$ & $\bar{F}\cdot\bar{u}_{\alpha}$ & $-k(\chi)(\sin\alpha(a+b\cos\beta)+b\cos\alpha\cos\beta\cos(\chi-\chi_0))$\\
  $F^{(b)}_{\alpha}$&$\bar{F}\cdot\bar{u}^{(b)}_{\alpha}$ & $k(\chi)(\sin\beta(a+c\cos\alpha)+c\sin\alpha\cos\beta\cos(\chi-\chi_0))$\\ \hline
\end{tabular} }
\tabnote{Superscript $b$ referes to definitions of the force components
at the other end of the $\chi$-rotatable bond. Factor $k$ includes the
Coulomb constant and the distance between the interacting parameter charges $q_1$ and
$q_2$. Parameters $(a,b,c,\alpha,\beta)$ are defined in Figure\ \ref{fig2} and
$\chi_0$ is a fixed angle of rotation value.}
\label{force}
\end{table}
The force components to the direction of the rotating bond, $\bar{F}_{\rho}$ and
$\bar{F}^{(b)}_{\rho}$,  do not contribute to the torque, because they are
parallel to the respective position vectors $\bar{c}$ and $\bar{b}$. By using
the force component definitions given in Table \ref{force}, the torque $\bar{M}$
in Equation\ \ref{Meq} is given as
\begin{align}
\bar{M}&=[cF_{\alpha}(\bar{u}_{\rho}\times\bar{u}_{\alpha}),cF_{\chi}(\bar{u}_{\rho}\times\bar{u}_{\chi}),bF^{(b)}_{\alpha}(\bar{u}^{(b)}_{\rho}\times\bar{u}^{(b)}_{\alpha}) ,bF^{(b)}_{\chi}(\bar{u}^{(b)}_{\rho}\times\bar{u}^{(b)}_{\chi})]\nonumber\\
& =[cF_{\chi}(-\bar{u}_{\alpha}),cF_{\alpha}\bar{u}_{\chi},bF^{(b)}_{\chi}(-\bar{u}^{(b)}_{\alpha}),bF^{(b)}_{\alpha}\bar{u}^{(b)}_{\chi}]
\label{Meq2}
\end{align}
In Equation\ \ref{Meq2}, rules for calculating cross products ($\times$) are used to
obtain the expressions on the second line. The moment of force $M_{\chi}$ can
be extracted from Equation\ \ref{Meq2} by representing the bond bending angle unit
vector as a sum of components:
$\bar{u}_{\alpha}=\sin\gamma\bar{u}_R+\cos\gamma\bar{k}$, $\gamma=\alpha,\beta$,
where $\bar{k}$ is the unit vector to the direction of the axis of rotation (the
z-axis in this case). Taking into account all simultaneous contacts $i$ and
reordering the terms to group those that are with respect to the end atoms and 
those that are with respect to the axis of rotation, the magnitude for the moment 
of force can be expressed as
\begin{align}
M&=|\sum_{i=1}^{N}M^{(i)}_{\alpha}|+|\sum_{i=1}^{N}M_{\alpha}^{(b,i)}|+|\sum_{i=1}^{N}M^{(i)}_{\chi}|+|\sum_{i=1}^{N}M^{(b,i)}_{\chi}|\nonumber\\
&=|\sum_{i=1}^{N}c_iF^{(i)}_{\alpha}|+|\sum_{i=1}^{N}b_iF^{(b,i)}_{\alpha}|+|\sum_{i=1}^{N}c_iF^{(i)}_{\chi}\sin\alpha_i|+|-\sum_{i=1}^{N}b_iF^{(b,i)}_{\chi}\sin\beta_i|\nonumber\\
&=|-\sum_{i=1}^{N}k_i(\chi)(c_i\sin\alpha_i(a_i+b_i\cos\beta_i)+b_ic_i\cos\alpha_i\sin\beta_i\cos(\chi-\chi^{(i)}_0))|\: +\nonumber\\
&\, +|\sum_{i=1}^{N}k_i(\chi)(b_i\sin\beta_i(a_i+c_i\cos\alpha_i)+b_ic_i\sin\alpha_i\cos\beta_i\cos(\chi-\chi^{(i)}_0))|\: +\nonumber\\
&+2\,|\sum_{i=1}^{N}k_i(\chi)b_ic_i\sin\alpha_i\sin\beta_i\sin(\chi-\chi^{(i)}_0)|
\label{SingleT}
\end{align}
In Equation\ \ref{SingleT}, the first two torque terms correspond to bond bending strain
$M_{\alpha}$, the third and fourth to torsional strain $M_{\chi}$. Subscript $b$
refers to that end of the rotatable bond which is parametrized with angle
$\beta$ and length $b$, as given in Figure\ \ref{fig2}. Vertical bars in Equation\
\ref{SingleT} mark an absolute value and $k(\chi)$ is given by Equation\ \ref{distSq}.
The bond angles $\alpha$ or $\beta$ ($\pi-\beta$), between a \textit{rotatable} bond and 
a \textit{rotating} ($\alpha$) or a reference bond ($\pi-\beta$) correspond, as
defined here, to the polar angle of the spherical polar coordinates in the fragment reference 
frame.\\

The x-axis of the fragment reference frame determines the zero point for 
$\chi$ and $\chi_0$ in Equations\ \ref{distSq} and \ref{SingleT}. Parameter 
$\chi_0^{(i)}$ (i=1,2,3) can have, for example, the values [$\frac{\pi}{3}$,$\pi$,$\frac{5\pi}{3}$]. 
The barriers of Equation\ \ref{SingleT} could be calculated using continuous charge 
distributions, but a simple point charge model is regarded as sufficiently 
representative of the distribution for our purposes here.\\

In practice, there are four independent force couples for each bond pair (four
bond charge and atom charge over-bond combinations, $q_{1,3}$ --- $q_{2,4}$) and
six or nine rotating covalent bond combinations. This means for barrier to
rotation calculations, that one uses a sum of 24 or 36 (parameter N) terms of
the form given in Equation\ \ref{SingleT}, with angle of rotation $\chi$ dependent
forces and position vectors. Note that in this work we use in practice only the
bond bending torque $M_{\alpha}$, which choise is discussed later in this section.\\

There is no straightforward connection between the functional forms of $M$ and a force 
field, as given in \cite{Mura2014,Heinz2013}, because the parameterizations are 
different, for example, $M$ combines both dihedral and bond bending angles 
in one function.
\subsection{Overlap method and interaction strength}
\subsubsection{Position distribution}
As described in this section, treatment of molecular contacts requires in our model, starting 
with a system of initial molecular structures, a distribution of allowed positions 
$\bar{r}_{\text{ref}}$. These are either positions of the atoms that are considered to interact 
with a studied fragment, or positions of interest with respect to a molecular fragment. The 
former is the so called Target-atom (NZ in Figure\ \ref{fig8}) and the latter needed in 
formulating the overlap integral, e.g., when the noncovalent bond is mediated with a small 
molecule like water. Mediated contacts were calculated in this way in ref \cite{Hakulinen2013}. 
It should be mentioned, that the roles (fragment side and Target side) can be reversed and 
naturally the same quantitative results should be obtained. The position distribution is 
here generated by rotations about relevant rotatable bonds in a structure, see 
Equations\ \ref{rs} and\ \ref{densityr} for details of the procedure used.\\

Positions found during a systematic mapping on a conformational space are chosen for further use 
based on level of torque in the related structural conformations. Form of the torque is a sum of 
the single-bond contributions given by Equation\ \ref{SingleT}. All conformations accessible on a 
thermal energy level are equally probable. The non-uniform distribution around a mean angle value 
in a rotamer library, see \cite{Harder2010} for a continuous library, appears when a distribution 
of levels is used. A single level can represent the mean thermal energy for one or several 
degrees of freedom, in which case the generated position distribution is an average, lacking more 
rare motions of larger amplitude but also the accrual, corresponding to small angular deviations 
from an equilibrium position (minimum torque).

\paragraph{Conformation generation} The relevant unit here is an amino acid residue, or 
more precisely, a lysine side chain. The 3D point distribution was generated from an 
initial Target-atom position $\bar{r}_{M}^{(0),(\text{ref})}$ 
with the following procedure:
\begin{equation}
\bar{r}^{(1->N)}_{\text{ref}}(\chi_{1},...,\chi_{n-2},\phi,\psi)=\sum\limits^{N}_{i=1}(\prod\limits_{j=N}^{i}R_{j}\,[\bar{r}_{M}^{(i-1),(\text{ref})}-\bar{r}_{M}^{(i),(\text{ref})}])+\bar{r}_{M}^{(N),(\text{ref})} 
\label{rs}
\end{equation}
In Equation\ \ref{rs}, the subscript $M$ refers to the Main-atom of a molecular fragment $f(i)$. The 
respective position vector together with rotations $R_i$ are determined separately for each step.
The rotations can all be with a different rotatable bond on the axis of rotation, all about the 
same bond, or something in between. Typically a fragment is involved more than once in the 
procedure (then at least two fragment definitions $f(i)$ and $f(j)$ are the same, i.e., defined 
by the same atom identifiers like NZ). Vector $\bar{r}^{(1->N)}_{\text{ref}}$ gives a single point 
produced by an arbitrary sequence of $N$ discrete rotation steps. The position of the reference 
point (starting from $\bar{r}_M^{(0),(\text{ref})}$) is tracked by ($n-2$) side chain angles of rotation 
and the overall conformation of the residue is additionally affected through two main chain angles of 
rotation.\\

In separating the main chain (mc) and side chain (sc) angles we presume that internal 
preferences on mc-side of $C_\alpha$ and on sc-side of $C_\alpha$ do not affect each other. This 
approximation is considered to give sufficient accuracy for our purposes here, though further 
development of the model will include mc torque parameters and a treatment of branch points. The 
dihedral angle indexing (1,...,n-2) in Equation\ \ref{rs} starts from the rotatable bond closest 
to the side chain terminal group, which is a convention to point out the order of varying the 
angles that is used here for a systematic coverage of conformational space.\\

Vector term $\bar{r}_{M}^{(N),(\text{ref})}$ is the Main-atom position of the last fragment used. The 
starting position vector $\bar{r}_{M}^{(0),(\text{ref})}$ that appears in the first term of the sum 
in Equation\ \ref{rs}, has also a more explicit definition that is given by Equation\ \ref{densityr}. 
The 3x3 rotation matrices $R_{i}$ combine orientations of the rotatable bond containing molecular 
fragment $f(i)$, with respect to the reference or base fragment frame (subscript ref), and a rotation 
in frame $f(i)$:
\begin{equation}
R_{i}(\Delta\chi_{i})=ROT_{(\text{ref}->M(i))}ROT_{\Delta\chi_{i}}ROT_{(M(i)->\text{ref})}
\label{rotation}
\end{equation}  
$\Delta\chi_i$ is the angular interval determining the rotation step in $f(i)$. Subscript $i$ indexes 
consecutive rotations. Position generation in Equation\ \ref{rs} includes visiting residues 
base fragment every time the rotatable bond changes, because the intermediate positions with 
respect to the base fragment are collected. An example set of side chain positions are 
represented in Figure\ \ref{fig7} by the blue patch-like clouds, which are a plotted 3D kernel 
density estimate, Equation \ref{targDensity}, of the generated Target-atom positions.

\subsubsection{Interaction model}
In the statistical model of either a Target-atom or a reference point (with respect to a 
fragment), information about the distribution is captured in a 3D probability density $f(\bar{r})$. 
The vector variable $\bar{r}$ can, as an example, be related to the starting position 
$\bar{r}_{M}^{(0),(\text{ref})}$ of a sequence of generated (Target-atom or reference point) positions, 
see Equation\ \ref{rs}, through
\begin{equation}
\bar{r}_k=ROT_{(\text{ref}->M(k))}\,\bar{r}+\bar{r}_{M(k)} ,\: \bar{r}_k=\bar{r}_{M}^{(0),(\text{ref})}
\label{densityr}
\end{equation}
Vector $\bar{r}_k$ in Equation\ \ref{densityr}, points to a chosen position with respect to the 
residues base fragment (e.g., $C_{\beta}$-$C_{\alpha}$-$C$ in lysine), subscript $\text{ref}$. In 
case $\bar{r}_k$ is the $k^\text{th}$ fragment Main-atom position vector $\bar{r}_{M(k)}$, the 
length of $\bar{r}$ equals zero. When the fragment, typically chosen for exploring a noncovalent 
contact, is the same as the residues base fragment, then $ROT_{(M(k)->\text{ref})}$ is the indentity 
matrix and $\bar{r}_k = \bar{r}+\bar{r}_{M}$. Combining these two cases gives origo, the base 
fragment Main-atom position.\\This conformational space managing scheme is primarily for one structural 
complex of the size involving, e.g., a few residues and ligands as flexible entities at a time. 
Incorporating larger scale structural motions is left for further study, as planned in Discussion.  
 
\paragraph{Overlap probability mass} Given a contact preference probability density for a 
fragment $f_p$ and a Target-atom density $g_t$, a quantitative measure for the strength of 
the contact is calculated as an overlap integral
\begin{equation}
OL=\int_{\mathbb{R}^3} f_{p}^{\frac{1}{2}}(\bar{r}(\bar{r}_k))\,g_{t}^{\frac{1}{2}}(\bar{r})\,\mathrm{d}^3\bar{r}
\label{OL}
\end{equation}
There $\bar{r}_k$ (as in Equation\ \ref{densityr}) is defined with respect to the reference frame of 
the studied fragment, as positioned and oriented in 3D space. The second equality in 
Equation\ \ref{densityr} means now that the present conformation, with corresponding position and 
orientation of the $k^{\text{th}}$ fragment, can be a starting structure for next structural 
manipulations. That is, probability density $f_p$ is defined relative to the frame of a studied 
fragment, but as seen from the base frame, where the target distribution $g_t$ is defined, or modelled.\\

The integral in Equation\ \ref{OL} is in principle over all space and the integrand is a 
preference--target overlap probability density. Depending on the definitions of $g_t$ and 
$\bar{r}_k$, the integrand corresponds either to a direct contact between structures (molecular 
fragment and a Target-atom or, in the future, two molecular fragments) or a bond mediated by at 
least one entity, like a water molecule or a metal ion. The value of the overlap integral includes 
the effect of structural flexibility, through the spatial form of the integrand. This is demonstrated 
in relation to the enthalpy-entropy compensation here in the corresponding section.\\

Determined by the plausible interactions, the overlap integral can contain $n>0$ preference densities. 
The generic equation with a target density $g_t$ is
\begin{equation}
OL_{\text{gen}}=\int_{\mathbb{R}^3}\sum\limits_{m=1}^{n}C_m\,f_{p,m}^{\frac{1}{2}}g_{t}^{\frac{1}{2}}\mathrm{d}^3\bar{r}
\label{OLgen}
\end{equation} 
which is, as Equation\ \ref{OL}, unitless in the sense that the outcome has dimensions of probability 
mass. The constant $C_m$ is introduced and has the meaning of an \textit{a priori} chemical or 
physical information on the relative non-directional preference of the interaction. For example, a 
distance and electronegativity based set of these prior weights was defined and used in our 
earlier work \cite{Hakulinen2012}. This scheme can be developed further, so that the individual weights 
express physical relations between contacts, which makes the overlap probabilities comparable, i.e., 
independent of the complex and fragment classification. Next we introduce a probabilistic way to treat 
the unfavorable contacts.

\paragraph{Negative probabilities as repulsion}
Contact preference densities $f_p$ in Equations\ \ref{OL} and \ref{OLgen} model \textit{attractive} 
nonlocal interactions. This follows from the molecular fragment classification and distance 
criteria imposed while collecting the training data (e.g., $\leq$ 3.3 \text{$\AA$}\ for a polar 
contact and $\leq$ 3.7 \text{$\AA$}\ for alleged dispersion \cite{Hakulinen2012}). It should be 
mentioned that, though the interacting fragments and Target-atoms are not free to move in a 
complex structural setting, the distance and direction preferences are considered representative 
of the realized attractive interactions in a large sample of observed structures. Modelled 
\textit{repulsion} needs to be added when structures and complexes are generated virtually, not 
observed experimentally.\\More precicely, after that conformations adhering only to internal 
constraints (i.e., torque) have been generated to study interactions, contacts to the molecular 
environment are taken into account separately. Starting from a set of molecular conformations 
that are internally, or locally, preferred and then incorporating the set of conformations in 
a complex, requires modeling repulsive external interactions in addition to the attractive ones. 
Repulsion is in this framework described as negative probability density. This approach is chosen to 
have probability as the one common measure for preference.

\paragraph{Incorporating negative probabilities} Negative probability density that overlaps with positive,
lowers the probability mass calculated for a contact. When the effect of repulsive components is 
surmounted by positive probability mass in the contact overall, formation of the complex can be
considered as possibly preferred. Then, in case the probability mass of the initial state is lower 
than in the suggested complex, the formation is taken as preferred. In terms of energy balance, 
negative overlap probability density (OPD) means positive total energy $E$, whereas positive OPD 
corresponds to negative $E$. The case in between these two, when there is no OPD, describes
$E=0$ and follows from no or only negligible interaction between the structural components.\\

The formal overlap calculation with, repulsion incorporating, contact preference probability density 
$f_{p,\pm}$ and target atom density $g_t$ is given here by
the equation
\begin{align}
OL&=\int_{\mathbb{R}^3}f_{p,\pm}^{\frac{1}{2}}g_t^{\frac{1}{2}}\mathrm{d}^3\bar{r} = \int_{\mathbb{R}^3}(\sqrt{f_{p,+}}+i\sqrt{f_{p,-}})\, \sqrt{g_t}\,\mathrm{d}^3\bar{r} = 
\label{probmass} \\
&=\int_{\mathbb{R}^3}(\sqrt{|f_{p,+}|}-\sqrt{|f_{p,-}|})\, \sqrt{g_t}\,\mathrm{d}^3\bar{r}=\int_{\mathbb{R}^3} (\sqrt{|f_{p,+}||g_t|}-\sqrt{|f_{p,-}||g_t|})\mathrm{d}^3\,\bar{r}
\nonumber
\end{align}
Subscript $+$ refers to a positive and $-$ to a negative probability density. Vertical bars again mark 
absolute values. The functional form of the integrand follows from that the negative and positive 
parts effectively vanish where the other one is nonzero. The negative part is placed in an imaginary 
component, so that before taking the absolute values they are not defined in the same probability space 
and are in this way represented as separate. The first term of the final integrand in the second 
line of Equation\ \ref{probmass} is the positive probability of the fragment--target interaction and the 
second term, as it is negative, lowers the value of the overlap probability mass $OL$. Normalization of 
the combined preference density $f_{p,\pm}$ in Equation\ \ref{probmass} is componentwise.\\

One can envision a three-body interaction to be represented as the integral of a function of the 
form $f^\frac{1}{3}g^\frac{1}{3}h^\frac{1}{3}$, but then the simple framework in Equation\ \ref{probmass} 
does not apply anymore. The question of more than two overlapping densities, should be studied in 
conjunction with developing the molecular fragment classification, which either explicitly or 
implicitly shoud have polarizability \cite{Gao2014,Mura2014} incorporated. Numerical examples in this work 
(section Numerical example) use an averaged probabilistic repulsive interaction model that has only 
distance dependence. It is normally distributed with standard deviation $\sigma$=0.85, which is between 
typical covalent and van der Waals radii. 

\paragraph{About the concept of negative probability} It is easy to find (in the internet) 
discussions on this topic by physicists and mathematicians \cite{Szekely2005}. In the context of 
physics, it can for example be seen as an intermediate technicality, such as in the Wigner 
probability function for coordinates and momenta \cite{Wigner1932}. In order to accept the concept 
as it is used here, one can consider a net negative overlap probability mass as representing an 
unstable contact, which can be stabilized through the rise of probability, for example as a more 
preferred contact is added to the interaction complex.

\subsubsection{Enthalpy--entropy compensation and overlap}
In the following, a linear relationship between enthalpy change and entropy change in 
complex formation -- a compensation effect -- is discussed for elemental contacts (one 
molecular fragment and one Target-atom). The contributions of elemental contacts can be 
combined in a straighforward way, by adding volumes and, possibly weighted, mean overlap 
densities. The differences in enthalpy and entropy change is in the compensation scheme 
here caused by a different contact type, like a modification to a ligand or a mutated amino 
acid residue. Also changes in solvent produce new types of elemental contacts, and therefore 
can be considered a part of the treatment. In contrast, to narrow down the treatment in 
this work, a change in temperature or a larger scale structural displacement, altering for 
example the structure of a binding site, can cause significant changes to Gibbs free energy 
of binding and therefore produce new conditions for the compensation to take place.\\

The phenomenon is relevant for a quantitative description of molecular interactions, and is 
analyzed, e.g., in articles \cite{Forrey2012,Sharp2001}. The motivation for a treatment here, is 
that it has not yet been decisively verified how extensive the enthalpy--entropy compensation effect 
is, as shown in the recent review \cite{Chodera2013}. The analysis in this work is related to the 
second topic (Conformational restriction) in the discussion on the physical origin of compensation 
in ref \cite{Chodera2013} and to Category 3 in ref \cite{Sharp2001}. Our probabilistic approach 
provides new tools for studying the mechanism behind compensation. Namely, in given conditions, the 
nature of the contact is described by the overlap probability density $\sqrt{f_p\,g_t}$, integrand 
of Equations\ \ref{OL} and\ \ref{probmass}. We use the concepts entropy $S$ and enthalpy 
$H$ already in relation to a single contact, where they correspond to freedom of motion and
constraints that limit the motion, respectively.\\

Technically, in case there is an enthalpy-entropy compensation effect between any two studied systems, the
ratio of the \textit{change in enthalpy change} ($\Delta(\Delta H)$) to the \textit{change in entropy 
change} ($\Delta(\Delta S)$) should be constant (T$_c$), i.e., not depend on $\Delta S$. This would produce 
a straight line for enthalpy change as a function of entropy change, $\Delta H=\Delta G + T_c\Delta S$, where 
the $\Delta H$-intercept is Gibbs free energy.\\About the formalism, assuming the numerical method chosen for solving the overlap integral converges to 
the correct OL, the choice of method, or implementation, should not influence the overall conclusions. So, in order to 
study here how enthalpy--entropy compensation might arise, using the same method as for calculations
(section Numerical example), the OL in Equations\ \ref{OL}-\ref{probmass} is represented as a Riemann sum
\begin{equation}
OL\approx\sum\limits^N_{i=1}(f_{p,\pm}^{\frac{1}{2}}g_t^{\frac{1}{2}})_iV_i=V\sum\limits^N_{i=1}(f_{p,\pm}^{\frac{1}{2}}g_t^{\frac{1}{2}})_i=V\sum\limits^N_{i=1}(h^{\frac{1}{2}}_{i,+}-h^{\frac{1}{2}}_{i,-})=V\sum\limits^N_{i=1}h^{\frac{1}{2}}_i
\label{RiemannOL}
\end{equation} 
In Equation\ \ref{RiemannOL}, $h_i^{\frac{1}{2}}$ is an abbreviated form of the density and $V_i$ are volume 
elements which all can be chosen to have equal size, $V_i=V$. This numerical integration can in principle 
be done with arbitrarily high precision (infinitely small value of $V$) inside the
\textit{overlap volume} $N\times V$, which is the size of the spatial area where overlap
density $h^\frac{1}{2}$ has values above some small threshold.

\paragraph{Entropy} In our probabilistic framework, entropy is considered
proportional to the opposite of the inverted overlap volume size:
\begin{equation}
S=-C_{OL}^{(S)} \frac{1}{NV},\: \:
[C_{OL}^{(S)}]=\frac{J}{K\AA^3}
\label{S}
\end{equation}
Here $N$ is the number of constant sized volume elements $V$ needed to cover the
overlap volume and $C_{OL}^{(S)}$ is coefficient of proportionality. This definition for 
$S$ is based on the rationale that, the contact that least decreases, or most increases, 
conformational freedom is entropically most favoured. And, this way entropy is directly 
tied to the properties of the overlap density. The quantification of $S$ can be 
considered to give the ratio of average volumetric overlap probability density to the 
overlap probability, which means that two contacts with different overlap mass (OL) from
the same overlap volume ($NV$) correspond to the same amount of entropy $S$. Also mentioned, 
that entropy $S$ is extensive with respect to the overlap volume $NV$, the larger $N$ is,
the less entropy is lost (or more obtained).

\paragraph{Enthalpy} A measure of enthalpy then, is given by how concentrated
the integrand $h^{\frac{1}{2}}=f_{p,\pm}^{\frac{1}{2}}g_t^{\frac{1}{2}}$ is,
i.e., how strongly peaked the probability density is, and accordingly, how
spatially limiting the bond is. This is represented as a coefficient times the
opposite of the overlap density mean (here, arithmetic mean of the $N$ Riemann 
sum integrand values),
\begin{equation}
H=-C_{OL}^{(H)}\langle h^{\frac{1}{2}}\rangle=-C_{OL}^{(H)}(\langle h^{\frac{1}{2}}_{+}\rangle-\langle h^{\frac{1}{2}}_{-}\rangle )
\label{H}
\end{equation} 
where the coefficient of proportionality $C_{OL}^{(H)}$ has the unit energy times 
volume, $J\AA^3$.\\Change in enthalpy change between two different contacts can be
expressed as $\Delta(\Delta H) = (H_{3}-H_{2})-(H_{1}-H_{0})$, where subscripts
0 and 2 refer to the situation before the studied contact is formed, for
example, fully hydrated state of an amino acid side chain. The same way, we
get for change in entropy change $\Delta (\Delta S) = (S_{3}-S_{2})-(S_{1}-S_{0})$.

\paragraph{Constant overlap and compensation} 
A constant overlap value for two realizations of a single type of interaction 
(a fragment class -- Target-atom class pair) corresponds to contacts of the same preference, 
i.e., with approximately uniform formation affinity from a similar initial state. Comparing 
two contacts of different type, requires also the \textit{a priori} weights $C_m$ in 
Equation\ \ref{OLgen}, but they are in the following included in the corresponding overlap 
probability masses (OL). Next, we show that a constant overlap (OL) in two contacts is the 
requirement for an exact compensation.\\

Two Riemann sums for the same OL (overlap), one of length $N_{i}$ and the other of length 
$N_{j}$, are equalized to get
\begin{equation}
N_iV\langle h_{i}^{\frac{1}{2}}\rangle=N_jV\langle h_{j}^{\frac{1}{2}}\rangle=OL
\label{dblsum}
\end{equation} 
It is important here that the two overlap volumes have different sizes, i.e., $N_{i}V \neq N_{j}V$, 
especially for even indeces, corresponding to final complexes. The different NV are relevant for 
that H-S compensation can occur, because otherwise entropy stays constant. The second equality 
in Equation\ \ref{dblsum} comes directly from the definition of the approximate overlap in 
Equation\ \ref{RiemannOL}. Using the conservation rule of Equation\ \ref{dblsum} we get for the 
change in entropy change (using also the definition of entropy in Equation\ \ref{S}),
\begin{align}
\nonumber 
&\Delta(\Delta S) = -C_{OL}^{(\text{S})}(\frac{1}{N_3V} - \frac{1}{N_2V}-\frac{1}{N_1V}+\frac{1}{N_0V}) =\\ 
&=-\frac{C_{OL}^{(\text{S})}}{V}(\frac{N_2N_1N_0-N_3N_1N_0-N_3N_2N_0+N_3N_2N_1}{N_3N_2N_1N_0})
\label{dds}
\end{align}
The change in enthalpy change is then, based on the definition in Equation\ \ref{H} and 
Equation\ \ref{dblsum},
\begin{align} 
\nonumber
&\Delta(\Delta H) = -C_{OL}^{(\text{H})}(\langle h_{(3)}^{\frac{1}{2}}\rangle-\langle h_{(2)}^{\frac{1}{2}}\rangle-\langle h_{(1)}^{\frac{1}{2}}\rangle+\langle h_{(0)}^{\frac{1}{2}}\rangle)= \\
&=-\frac{C_{OL}^{(\text{S})}}{V}(\frac{N_2N_1N_0OL_3-N_3N_1N_0OL_2-N_3N_2N_0OL_1+N_3N_2N_1OL_0}{N_3N_2N_1N_0})
\label{ddh}
\end{align}
Dividing $\Delta(\Delta H)$ in Equation\ \ref{ddh} with $\Delta(\Delta S)$ in Equation\ \ref{dds}, we 
get 
\begin{align}
\nonumber
&\frac{\Delta (\Delta H)}{\Delta (\Delta S)}=\frac{C_{OL}^{(\text{H})}}{C_{OL}^{(\text{S})}}(\frac{N_2N_1N_0OL_3-N_3N_1N_0OL_2-N_3N_2N_0OL_1+N_3N_2N_1OL_0}{N_2N_1N_0-N_3N_1N_0-N_3N_2N_0+N_3N_2N_1})=\\
&=\frac{C_{OL}^{(\text{H})}}{C_{OL}^{(\text{S})}}OL\,;\: \: OL=OL_i, \: \: i=1,2,3,4
\label{EEc}
\end{align}
Equation\ \ref{EEc} shows how a constant OL produces a constant $\Delta(\Delta H)$ to 
$\Delta(\Delta S)$ ratio, corresponding to the \emph{compensation temperature} $T_{c}=\frac{C_{OL}^{(\text{H})}}{C_{OL}^{(\text{S})}}OL$. The curve $\Delta H (\Delta S)$ is a straight line over the range where OL is constant, which corresponds to \textit{strong} compensation. This demand for a constant overlap probability mass for both initial and final states of both binding processes, is a strict condition. It would mean equal preferences for all four states and therefore a nonspontaneous binding process, due to the lack of a driving force. A more realistic situation could be that, the initial states (indeces 0 and 2) are similar enough that they can be approximated to cancel out in $\Delta (\Delta S)$ and $\Delta (\Delta H)$. In Equation\ \ref{EEc} this means that N$_0$=N$_2$=1, OL$_0$=OL$_2$=0 and OL$_1$=OL$_3$=OL. This result, on one possible source of compensation, suggests the same conclusion as in review \cite{Chodera2013}, that a \textit{weak} form of compensation is more likely to be real than the \textit{strong}, or nearly exact.\\

Still about the coefficients of proportionality, the enthalpy related $C_{OL}^{(H)}$ is inferred to be positive, since then the enthalpy change $\Delta H < 0$ when
\begin{equation}
 \langle h_{(1)}^{\frac{1}{2}}\rangle-\langle h_{(0)}^{\frac{1}{2}}\rangle > 0
\end{equation} 
The coefficient relating to entropy ($C_{OL}^{(S)}$) is defined as positive, though a situation where the overall OL is negative (i.e., repulsion is stronger the attaction and the system is unstable) could require a negative constant, which is not studied here further.    
\paragraph{Ensemble level} A direct formal link between the result in Equation\ \ref{EEc} and a 
statistical ensemble is represented with the aid of the single contact overlap integral and 
change in Gibbs free energy change, as expected value in an ensemble:
\begin{align}
&OL(E_i)=\int_{\mathbb{R}^3} f_{p,\pm}^{\frac{1}{2}}(\bar{r})g_{t}^{\frac{1}{2}}(\bar{r};E_i)\,\mathrm{d}^3\bar{r} \: \rightarrow \: p_i = p(E_i)=p(OL_i) 
\nonumber \\
&\Delta(\Delta G)=\sum_{i=1}^np_i[\Delta(\Delta H_i)-T\Delta(\Delta S_i)]\approx 0,\: T\approx T_c =\frac{C_{OL}^{(\text{S})}}{C_{OL}^{(\text{H})}}OL
\label{ensembleG}
\end{align}
Here $n$ is the number of complexes (contacts) in the ensemble, $T_c$ refers to the compensation 
temperature and $p_i$ represents the probability of a thermal energy level $E_i$, which is a 
parameter for the target atom distribution $g_t$. A thermal energy dependence is also true for 
the fragment preference density $f_{p,\pm}$, but because overlap then still has the same probability 
as the energy state has, we use a fixed $f_{p,\pm}$ for clarity. The second line in 
Equation\ \ref{ensembleG}shows the ensemble level (each i separately) for enthalpy--entropy 
compensation, which is exact when the absolute temperature  $T=T_c=C_{OL}^{(H)}/C_{OL}^{(S)}OL$.\\

A molecular contact that has many simultanenous interactions and several levels of motion 
\cite{Andrusier2008} involved, likely gets an approximate compensation effect at best. As suggested, 
this can be studied by using the overlap probability densities in varying molecular settings. We 
conclude this topic by noting that also processes such as hydration are ultimately forming and breaking 
of molecular contacts -- a balance between freedom of motion and strength of interaction -- so the 
basic reasoning presented here can be applied generally.

\subsubsection{Probability density functions used}
The parametric contact preference density describing attractive interactions is formulated 
as a mixture of one-dimensional densities:
\begin{equation}
f_{p}(\rho,\theta,\phi)=\sum\limits_{i=1}^{n}\text{N}(\rho;\hat{\mu}_i,\hat{\sigma}_i^2) \,\text{vM}(\theta;\hat{\mu}_i^{(\theta)},\hat{\kappa}_{i})\,[\sum\limits_{j=1}^{n_{j}}\text{N}(\phi;\hat{\mu}_{ij},\hat{\sigma}_{ij}^2)]
\label{prefDensity}
\end{equation}
In Equation\ \ref{prefDensity}, function N refers to the normal distribution and vM
to von Mises distribution. Parameters $(\hat{\mu}_i,\hat{\sigma}_i^2,\hat{\mu}_i^{(\theta)},...)$ 
get values as Bayesian estimates obtained using Protein Data Bank structures, and have so far been 
maximum \textit{a posteriori} estimates or posterior modes\cite[pp.~37-38]{Gelman2004}. The density 
in Equation\ \ref{prefDensity} was introduced in a previous study \cite{Hakulinen2012}, where the chosen 
functional form followed from Kolmogorov-Smirnov normality tests, among others. Also, the spatial 
distribution of the modeled atom positions depends on the molecular fragment classification and the 
amount of data available for model training, therefore, some degree of exploratory character is still preserved in the statistical model. It is achieved for the 3D probability distribution by using 
interconnected 1D densities as shown in Equation\ \ref{prefDensity}, because this way not too much 
regularity in the target distributions is assumed. The form of the distribution is able to adapt to 
new data for modest computational cost in comparison with, for example, a kernel estimate, through 
using some simplifying assumptions and therefore fewer terms. This \textit{nonlocal} contact part of 
the model bears a resemblance with a traditional force field \cite{Mura2014}, due to the partly 
predetermined functional form, but a quantum mechanics based approach, similar to the method used 
in ref \cite{Gao2014}, could also be possible.\\

For this work, the probabilistic repulsion density (see Equation\ \ref{probmass}) was modelled as 
a distance $\rho$ dependent normal distribution 
\begin{equation}
\sqrt{|f_{p,-}||g_t|}=\frac{1}{(2\pi\sigma^2_{\text{rep}})^{\frac{3}{2}}}\exp(-\frac{\rho^2}{2\sigma_{\text{rep}}^2})
\label{repN}
\end{equation}
The value used for variance, given also in Figure\ \ref{fig10}, was 
$\sigma^2_{\text{rep}}=0.7225$ {\AA\textsuperscript{2}}. This value, which corresponds to standard 
deviation that is between typical estimate intervals for covalent and van der Waals radii, was also 
chosen because using that in Equation\ \ref{repN} produced a reasonable steric interaction map for a 
lysine residue, see upper left plot in Figure\ \ref{fig10}. Repulsion in Equation\ \ref{repN} was 
applied to all plausible noncovalent contacts in the structure (actually for the dipeptide, but the 
result is considered to represent that of a residue).\\ 

The function $g_t$ in the overlap integrand in Equation\ \ref{OL} models a target atom distribution, 
and was in this work modeled with a 3D kernel density estimate:
\begin{equation}
g_{t}(\bar{r})=\frac{1}{N_t} \sum\limits_{i=1}^{N_{t}}\frac{\exp(-\frac{1}{2}(\bar{r}-\bar{\mu}_i)^T\mathbf{H}_t^{-1}(\bar{r}-\bar{\mu}_i))}{(2\pi)^{\frac{3}{2}} \sqrt{|\mathbf{H}_t|}}
\label{targDensity}
\end{equation}
where $\mathbf{H}_t$ is the bandwidth matrix, $\mathbf{H}_t^{-1}$ its inverse matrix and $|\mathbf{H}_t|$ 
its determinant. $\mathbf{H}_t$ was determined using statistical modeling environment R 
package 'ks'\cite{Duong2014}, on sets of target atom positions obtained using a torque model for 
the internal rotations, as described in section Electrostatic barrier to rotation (see 
Equations\ref{SingleT} and \ref{rs}). A more simple form with less terms for $g_t$ would lower the 
computational cost of the overlap calculation. This could be found by using the kernel density estimate 
\ref{targDensity} as a starting point.
\paragraph{Model layers}
Before we give numerical examples from applying the method, the three layers of the 
modelling scheme are briefly discussed. The first two layers are overlap probability mass
{m\textsuperscript{p}} for molecular fragment contact preference densities (incorporating 
probabilistically modelled repulsion) with statistically modelled target distributions, and 
the internal torque $M$ of the structure. The sum of the products of overlap masses with 
torque weights ($\text{w}^{(M)}$) is still in general weighted with fragment class specific 
\textit{a priori} probabilities $C_i$ (Equation\ \ref{OLgen}) for each contact type:
\begin{equation}
\text{m}^{(\text{p})} \rightarrow \text{w}^{(M)}_j \times\text{m}^{(\text{p})}_j\rightarrow \text{C}_i
\sum_j \text{w}^{(M)}_{ji}\times \text{m}^{(\text{p})}_{ji}
\end{equation}       
Subscript $j$ indexes conformations, which can also refer to a continuous variable, and $i$ 
refers to molecular fragment class--contact class pairs. The $\text{w}^{(M)}_{ji}$ term represents 
conformation-specific weights based on structural torque and is here needed only for the molecular 
fragment side $f_p$, because the internal preferences are represented by a distribution ($g_t$) 
that is based on torque. Similar weighting can in principle also be applied to $f_p$, in which 
case the second layer vanishes and the hierarchy reduces to 
m$^{(\text{p})}\rightarrow$ C$_i$m$^{(\text{p})}_i$. However, the topic of this work 
is to deal with the first level {m\textsuperscript{p}}.

\section{Numerical Results}

\subsection{Butane as training target}
The n-butane molecule was used as a model system to obtain torque parameters for 
generating lysine side chain conformations, needed in the numerical example. A butane 
rotation barrier derived from experimental measurements (infrared spectroscopy) 
\cite{Herrebout1995,Mo2010} was used as a reference when estimating the parameters, given in 
Table \ref{BTNparams}. The local maxima, staggered conformations, received the target value 
3.62 kcal/mol and the gauche local minima 0.67 kcal/mol, over the minimum torque 
(\emph{trans}-conformation).\\

The background torque in these calculations, was based solely on the bond bending part M$_{\alpha}$ 
of the torque in Equation\ \ref{SingleT}. The reason for this is that the contribution of 
M$_{\chi}$ towards the barrier profile remained unclear to us (see Figure\ \ref{fig3}), whereas 
$M_{\alpha}$ has the typical form of butane barrier to rotation, as found in literature. On the 
other hand, $M_{\chi}$ changes more rapidly and contributes neither in staggered nor eclipsed 
conformations, but only in the angle intervals between them. Since we are unsure if these 
intervals influence the information given by the experimental method, the traditional form given 
by $M_{\alpha}$ will be used.
\subsubsection{Parameter estimation}
\begin{figure}[ht]
\includegraphics{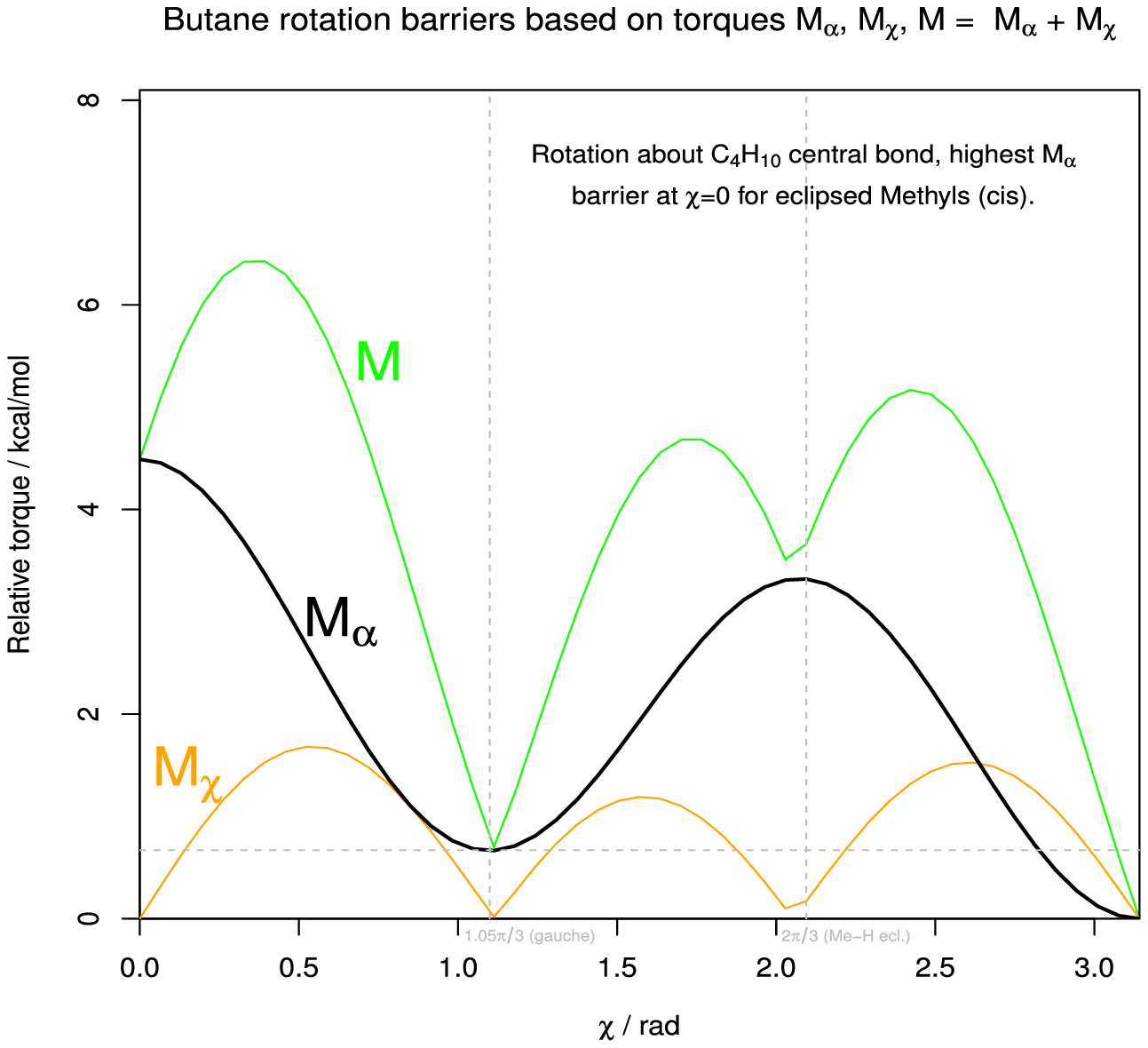}
\caption{Butane torque barrier to rotation. Black curve with label M$_{\alpha}$ is the bond 
bending torque used for numerical examples in this work. Orange curve with label M$_{\chi}$ is 
torsional torque and green with label M is the sum M$_{\alpha}$+M$_{\chi}$. Vertical dashed lines 
mark reference angle values $(1.05*\pi/3,2\pi/3)$. The zero point (radians) corresponds to 
\emph{cis}, $\chi=1.05*\pi/3$ to a staggered conformation producing a local minimum 
and $\chi=2\pi/3$ to an eclipsed conformation with a local peak. The horizontal dashed gray 
line corresponds to an experimental measurement of the gache$_+$ minimum at $1.05*\pi/3$, see text 
for reference. The angle value $\chi=\pi$ (\emph{anti} or \emph{trans}) has the lowest torque, 
which was defined as zero level. See Figure\ \ref{fig2} for a definition of the torque components.}
\label{fig3}
\end{figure}

The torque parameters were estimated numerically with 
the criteria that the calculated bond bending torque curve (M$_{\alpha}$) matches accurately 
the directly measured gauche$_\pm$ value 0.67 kcal/mol and then gets values as close as possible 
to those deduced in the experimental work reporting article \cite{Herrebout1995} for cis-conformation 
and local maxima ($\pm2\pi/3$). The experimental reference value is indicated with a horizontal 
gray dashed line in Figure\ \ref{fig3}. The vertical dashed lines mark the observed gauche$_+$ 
minimum position at $1.05*\pi/3$ \cite{Herrebout1995} and local maximum ($2\pi/3$) between gauche$_+$ 
and anti- or trans-conformation. The $M_{\alpha}$ barrier profile calculated is most reminiscent of 
the one produced with the block-localized wave function (BLW) approximation, figure 1 in ref 
\cite{Mo2010}, which does not include the effect of hyperconjugation. Our torque model here has 
a somewhat lower \emph{cis}-value and a somewhat higher $2\pi/3$-value than the BLW-curve, therefore 
being closer to the curve derived relating to the infrared study \cite{Herrebout1995}. Both, our 
torque model and BLW get the gauche$_+$ minimum practically at the mentioned, experimentally determined 
angle of rotation value $\chi=1.05*\pi/3$ (or 62.8$^o$ as given in the article).\\
\subsubsection{Related barriers to rotation}
\begin{figure}[ht]
\includegraphics{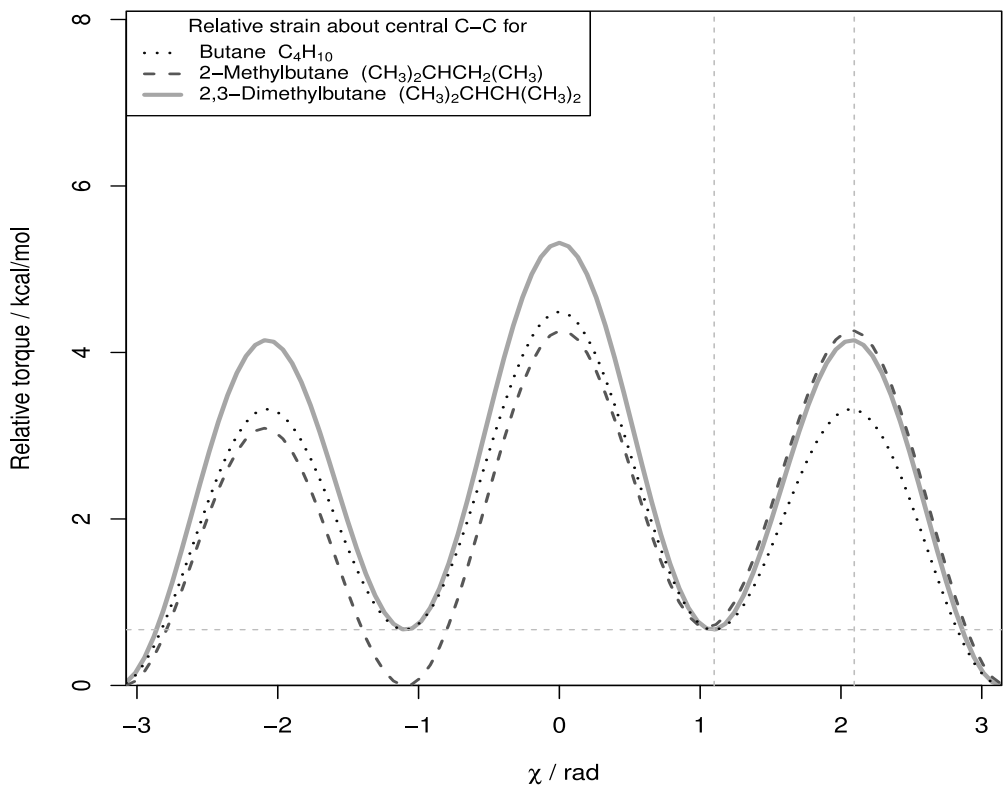}
\caption{Calculated rotational barriers of methylated butanes, compared with that of n-butane. 
Structures were rotated internally about the central carbon-carbon bond and strain was calculated 
as bond bending torque (M$_{\alpha}$). The strain at gauche$_+$ has the same value (0.67 kcal/mol) 
for all the three molecules. Torque at the other local minimum, gauche$_-$ ($-2\pi/3$), is for 
2-methylbutane the same as at trans-conformation, i.e., zero because the lowest torque is used 
as the base value. 2-Methylbutane has only one staggered conformation (gauche$_+$) with higher than 
minimum torque, due to asymmetry in groups over at the rotatable bond (two and one methyls bonded 
to the end atoms). Lowest strain for 2,3-methylbutane, i.e., \emph{trans}-conformation or 
\emph{anti}, is when the two hydrogens bonded to C$_2$ and C$_3$ are \emph{trans}. The set of 
parameters given in text were used.} 
\label{fig4}
\end{figure}

The set of parameters given in Table \ref{BTNparams} were used to produce also other than 
the butane rotational barrier, to show that these tentative point charge parameters can be 
used more widely and get coherent results. In Figure\ \ref{fig4}, we show a comparison of the 
butane curve to two methyl-substituted butanes (2-methylbutane and 2,3-dimethylbutane). The 
gray dashed lines are the same as in Figure\ \ref{fig3} and the curves approximately coinside 
at gauche$_+$, but 2-methylbutane gets a lower value at gauche$_-$ ($\chi\approx -\pi/3$). In 
the torque model, this follows structurally from that 2-methylbutane has two staggered conformations 
without two instances of three consecutive Methyls or Hydrogens (as seen in a Newman projection), 
whereas butane and 2,3-methylbutane have only one such conformation (methyls or hydrogens in 
\emph{trans}-conformation, respectively). 
\paragraph{Barriers in light of rules from organic chemistry}
The form of the torque barriers can be verified against textbook organic chemistry, e.g., 
the rules for estimating strain given in ref \cite[p.~161]{Klein2010}. Using those - 11 kJ/mol 
for eclipsed and 3.8 kJ/mol for gauche methyls (Me), 4 kJ/mol for H$\leftrightarrow$H and 
6 kJ/mol for H$\leftrightarrow$Me eclipsed - produces the same functional forms for 
2-methylbutane and 2,3-dimethylbutane as given in Figure\ \ref{fig4} here.\\ 

In case of 2-methylbutane, the peak heights are quite close to those estimated with the 
mentioned directional rules (our 4.3 vs. 4.1 kcal/mol and our 3.0 vs. 3.4 kcal/mol). For 
2,3-methylbutane the same are 5.3 vs. 4.4 kcal/mol and 4.1 vs. 3.7 kcal/mol, somewhat 
more different values. The joined gauche minimum torque for all three in Figure\ \ref{fig4} 
is also produced by the strain rules \cite[p.~161]{Klein2010}, but with a higher value of 
0.9 kcal/mol as compared with the experimental value 0.67 kcal/mol form ref \cite{Herrebout1995}. 
As mentioned, we used the latter as the most reliable fixed point for fitting the butane 
barrier to rotation. 
\subsubsection{Four ethane-type rotational curves}
\begin{figure}[ht]
\includegraphics{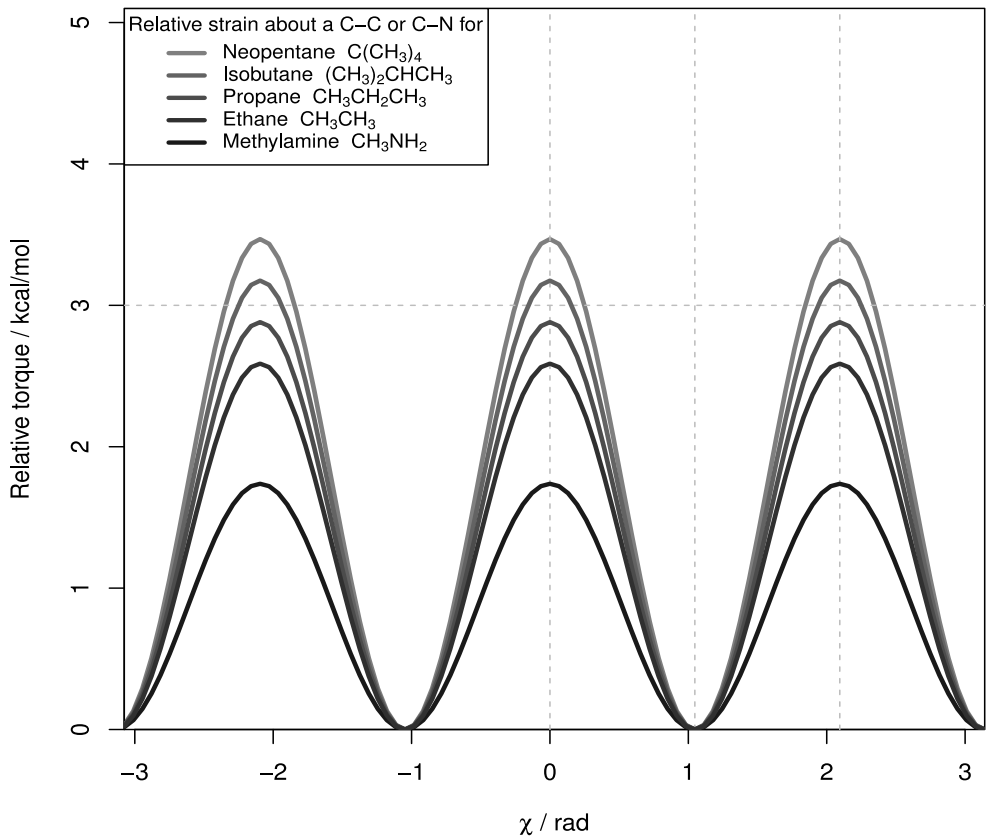}
\caption{Bond bending torque rotational barrier for ethane and for internal rotations that have 
similar functional form of barrier in four other molecules. The molecules are rotated internally 
with respect to a methyl group in the structure. The figure shows how, in this model, a 
carbon-carbon (single) bond rotating at the other end of a rotatatable bond, experiences larger 
torque than a carbon-hydrogen bond. This then produces the result of higher barrier for eclipsed 
methyl group and hydrogen, than for eclipsed hydrogens. The set of parameters given in 
Table \ref{BTNparams} were used, except an adjusted set for methylamine (lowest barrier) to obtain 
expected barrier height, which adjustment most likely produces a real difference due to the 
influence of the nitrogen (see text for details).}
\label{fig5}
\end{figure}

In Figure\ \ref{fig5} is plotted ethane rotational curve with four other having the same 
general form, but different heights. In each of the four reference molecules, there is a 
methyl group at the other end of the rotatable bond studied, which produces the ethane-like 
torque profile. The strength is varied by groups bonded to the other end of the rotatable bond, 
and is highest for neopentane with three methyls bonded there. It is noted that the torque model, 
at least in the present form, handles a methyl as a carbon, which means specifying the covalent 
bond length and positions of parameter charges differently than, e.g., for a hydrogen at the 
other end of the rotating bond (see Table\ \ref{BTNparams}).\\

The lowest barriers are for methylamine, for which an adjustment to the parameters were made 
due to the amino groups, nitrogen having different electronegativity than carbon and (here) 
two hydrogens instead of the three in a methyl group. The adjustment was targeted to produce 
the barrier height that is about 2/3 of the ethane barrier calculated with the parameters 
in Table \ref{BTNparams}, 2.55 kcal/mol, to demonstrate applying the model.
\subsubsection{Comparison with a quantum chemical two-diheral map}
\begin{figure}[ht]
\includegraphics{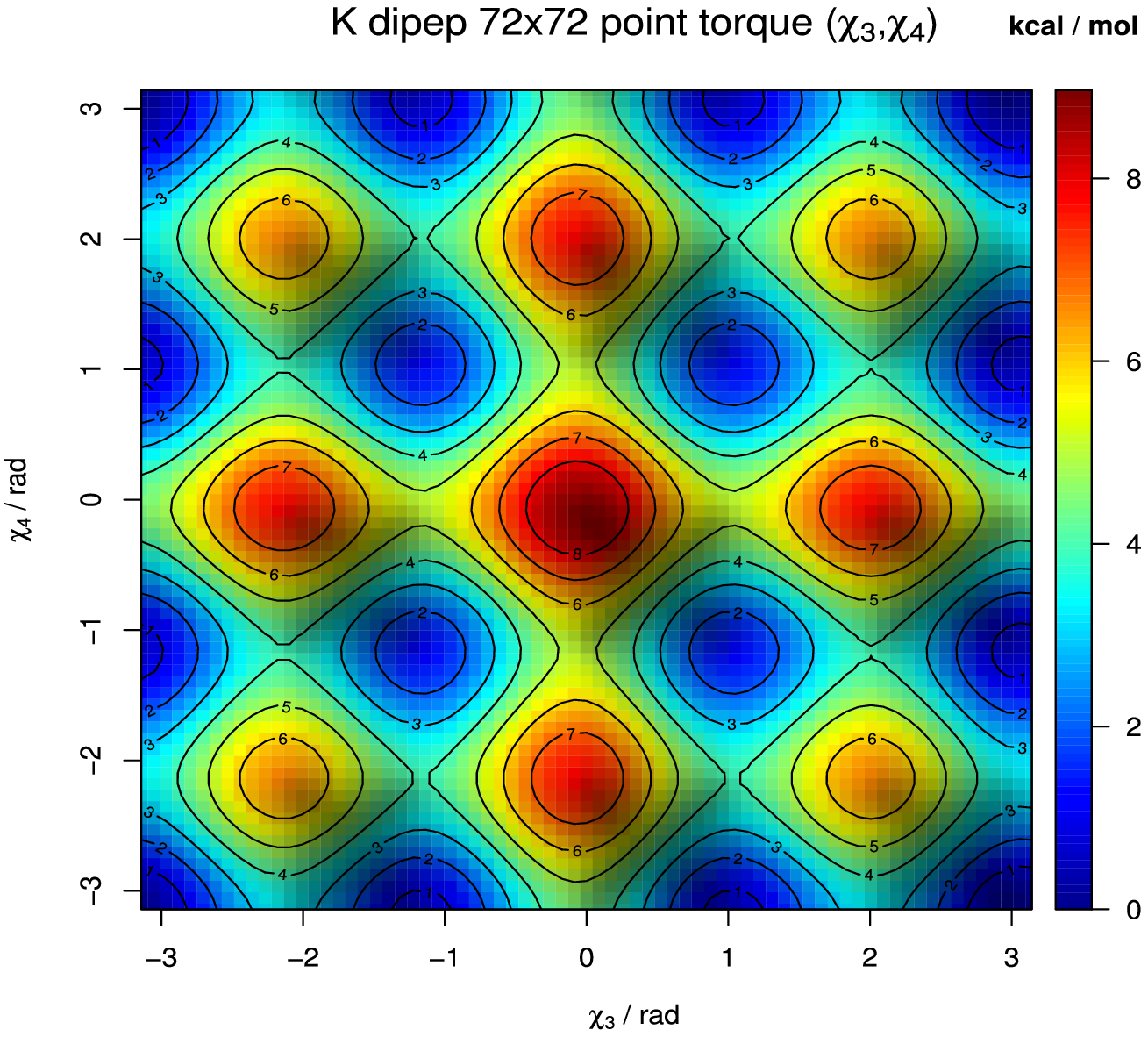}
\caption{Torque map for angles ($\chi_3$,$\chi_4$) in lysine side chain, calculated as bond bending 
torque M$_{\alpha}$. The corners correspond to \emph{trans}-conformation for both dihedrals, about 
C$_{\gamma}$-C$_{\delta}$ and C$_{\delta}$-C$_{\epsilon}$. }
\label{fig6} 
\end{figure}

The third test for this set of parameters, in Table \ref{BTNparams}, is that they were used 
to produce the map for two consecutive straight chain alkane angles of rotation, representing 
the dihedrals $\chi_3$ and $\chi_4$ in lysine side chain.The result in Figure\ \ref{fig6} 
was (visually) compared with a quantum chemically calculated map for two central diherals of 
n-pentane in ref \cite{Martin2013}. Our torque map is both qualitatively and quantitatively a 
good match to the quantum chemically calculated torsion surface in figure 1 of article 
\cite{Martin2013}. The only feature not observed in Figure\ \ref{fig6} here, is the so called 
pentane interference, which is an interaction between parts of the alkane more distant than 
three bonds apart. In our framework, such interactions belong to external contacts (torque 
being defined as the internal or local).\\
\begin{table}
\tbl{Torque parameters used for calculating internal rotation preferences with respect 
to dihedrals $\chi_2$ - $\chi_4$ in lysine side chain.}
{\begin{tabular}{| l | r | r | r | r | r |} \hline
 b / \AA  & $\beta$ / degs. & c / \AA & $\alpha$ / degs. & qprod / e$^2$ & $\chi_0$/degs.\\
\hline 
  0.7650 & 70.5 & 0.7650 & 70.5  & 3.31 & -180\\
\hline 
  1.5300 & 70.5 & 1.5300 & 70.5  & 0.37 & -180\\
\hline 
  0.7650 & 70.5 & 1.5300 & 70.5  & -1.11 & -180\\
\hline 
  1.5300 & 70.5 & 0.7650 & 70.5  & -1.11 & -180\\
\hline
  0.7650 & 70.5 & 0.6867 & 70.5 & 3.31 & -180\\
  \hline
  1.5300 & 70.5 & 1.0900 & 70.5 & 0.34 & -180\\
\hline 
  0.7650 & 70.5 & 1.0900 & 70.5 & -1.00 & -180\\
\hline 
  1.5300 & 70.5 & 0.6867 & 70.5 & -1.11 & -180\\
\hline 
  0.7650 & 70.5 & 0.6867 & 70.5 & 3.31 & -180\\
  \hline 
  1.5300 & 70.5 & 1.0900 & 70.5 & 0.34 & -180\\
 \hline 
  0.7650 & 70.5 & 1.0900 & 70.5 & -1.00 & -180\\
\hline 
  1.5300 & 70.5 & 0.6867 & 70.5 & -1.11 & -180\\
\hline 
  0.6867 & 70.5 & 0.7650 & 70.5 & 3.31 & -60\\
\hline 
  1.0900 & 70.5 & 1.5300 & 70.5 & 0.34 & -60\\
  \hline 
  0.6867 & 70.5 & 1.5300 & 70.5 & -1.11 & -60\\
\hline 
  1.0900 & 70.5 & 0.7650 & 70.5 & -1.00 & -60\\
  \hline 
  0.6867 & 70.5 & 0.6867 & 70.5 & 3.31 & -60\\
\hline 
  1.0900 & 70.5 & 1.0900 & 70.5 & 0.30 & -60\\
  \hline 
  0.6867 & 70.5 & 1.0900 & 70.5 & -1.00 & -60\\
\hline 
  1.0900 & 70.5 & 0.6867 & 70.5 & -1.00 & -60\\
  \hline 
  0.6867 & 70.5 & 0.6867 & 70.5 & 3.31 & -60\\
\hline 
  1.0900 & 70.5 & 1.0900 & 70.5 & 0.30 & -60\\
  \hline 
  0.6867 & 70.5 & 1.0900 & 70.5 & -1.00 & -60\\
\hline 
  1.0900 & 70.5 & 0.6867 & 70.5 & -1.00 & -60\\
  \hline 
  0.6867 & 70.5 & 0.7650 & 70.5 & 3.31 & 60\\
\hline 
  1.0900 & 70.5 & 1.5300 & 70.5 & 0.34 & 60\\
  \hline 
  0.6867 & 70.5 & 1.5300 & 70.5 & -1.11 & 60\\
\hline 
  1.0900 & 70.5 & 0.7650 & 70.5 & -1.00 & 60\\
  \hline 
  0.6867 & 70.5 & 0.6867 & 70.5 & 3.31 & 60\\
\hline 
  1.0900 & 70.5 & 1.0900 & 70.5 & 0.30 & 60\\
  \hline 
  0.6867 & 70.5 & 1.0900 & 70.5 & -1.00 & 60\\
\hline 
  1.0900 & 70.5 & 0.6867 & 70.5 & -1.00 & 60\\
  \hline   
  0.6867 & 70.5 & 0.6867 & 70.5 & 3.31 & 60\\
\hline
  1.0900 & 70.5 & 1.0900 & 70.5 & 0.30 & 60\\
  \hline   
  0.6867 & 70.5 & 1.0900 & 70.5 & -1.00 & 60\\
\hline
  1.0900 & 70.5 & 0.6867 & 70.5 & -1.00 & 60\\
  \hline 
\end{tabular}}
\tabnote{Rotation about C$_\alpha$-C$_\beta$ ($\chi_1$) had a somewhat modified set due to not 
being alkane bond like. Model structure for estimating these parameters was n-butane. The 
tabulated values are to be taken strictly as suitable parameters for the torque model, not as 
estimates of, for example, the mean bond angle values. See Figure\ \ref{fig2} for parameter 
definitions and text for details of estimating the parameters.}
\label{BTNparams}
\end{table}

\subsubsection{Integration details} Target-atom distributions were in this work systematically 
generated through internal rotations about single bonds and modelled with a 3D kernel density 
estimate (see Equation\ \ref{targDensity}). The bandwidth matrix $\mathbf{H}_t$ was evaluated 
for each thermal energy level (equals to torque cutoff) separately, using the default plug-in 
selector of the statistical modeling environment R package 'ks' \cite{Duong2014}. During the 
Riemann sum overlap calculations, the finite volume element 
$\Delta V = \frac{1}{3}(\rho_{i+1}^3-\rho_i^3)(\cos(\theta_{j})-\cos(\theta_{j+1}))(\phi_{k+1}-\phi_{k})$ 
is kept constant by generating the variable values as the following sequences:
\begin{align}
&\rho_i = (i\,\frac{4}{3}\pi)^{\frac{1}{3}}\rho_1, \: i=1,...,n_{\rho}
\nonumber\\
&\theta_j = \arccos(1-j+j\cos(\theta_1)), \: j=1,...,n_{\theta}\\
&\phi_k = k\,\phi_1, \: k=1,...,n_{\phi}
\nonumber
\end{align} 
The values for parameters $\rho_1,\theta_1, \phi_1 $ is a set of chosen starting values that 
determine the size of $\Delta V$.

\subsection{Example: Conformational preferences in a lysine dipeptide}
\begin{figure}[ht]
\includegraphics{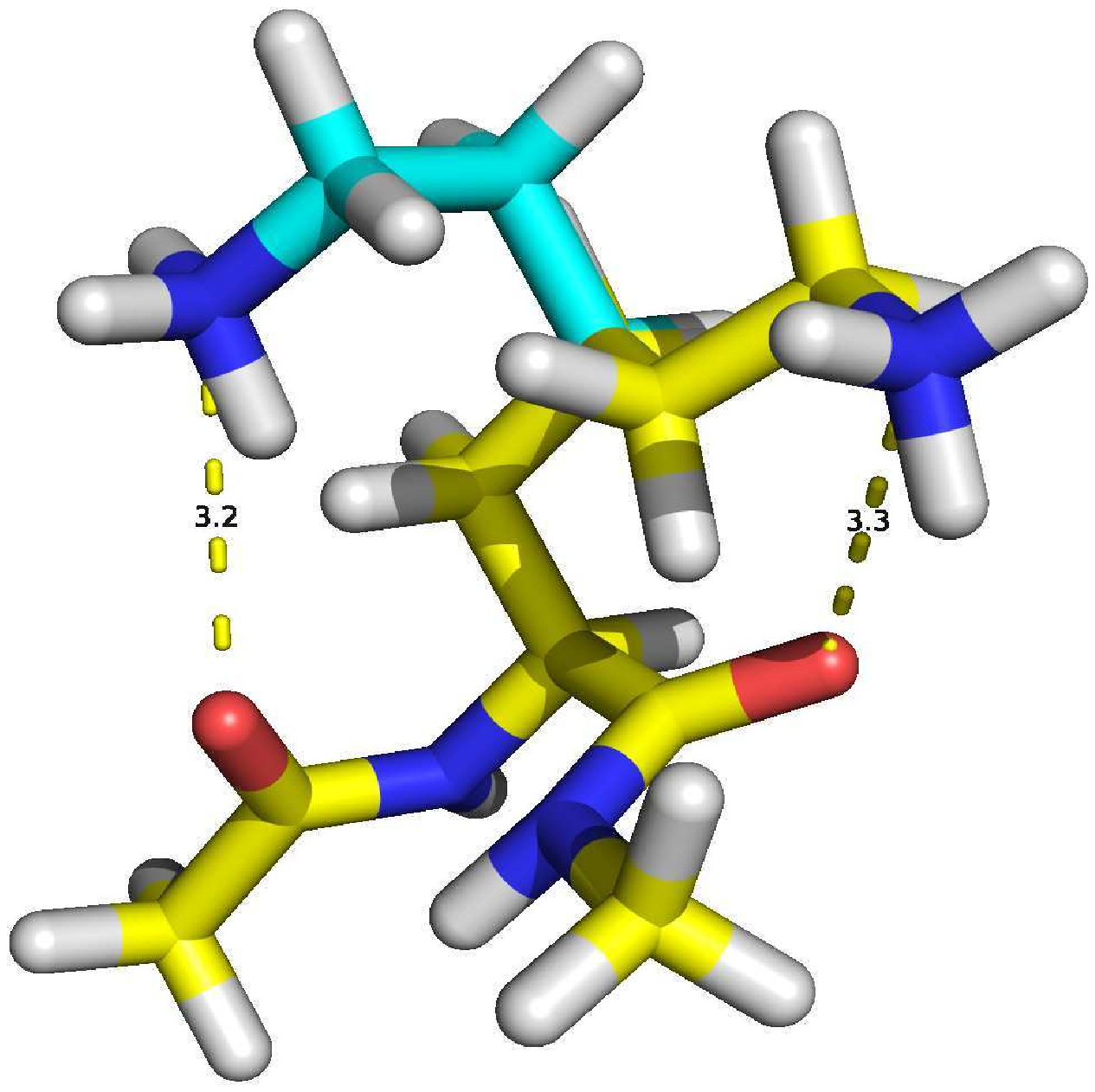}
\caption{A lysine dipeptide model system. Main chain (mc) is for
both contacts in a conformation corresponding to modest overlap. The angle 
values are ($\phi$,$\psi$)=(75,-30) degrees. This mc conformation is not necessarily
internally favoured (see text for details), but is used to visualize the
contact. Shown are also two side chain (sc) conformations that correspond to
weak hydrogen bonds between sc amino nitrogen (NZ) and mc carbonyl oxygen. The
side chain conformations experience internal torque that is less than 1.2 kcal/mol.}
\label{fig7} 
\end{figure}

We have here as an exemplifying case a N-acetylated and {N\textsuperscript{'}}-methylamidated 
lysine dipeptide structure, shown in Figure\ \ref{fig7} in two distinct side chain conformations. The 
dipeptide model structure was adapted from the work by Zhu \emph{et al.} \cite{Zhu2012}, where it was 
used as one of the targets for a quantum mechanical investigation of internal energy landscapes. Here, 
we use more degrees of freedom than in the original work and, in addition, study one internal
noncovalent contact type in detail.\\

\begin{figure}[ht]
\includegraphics{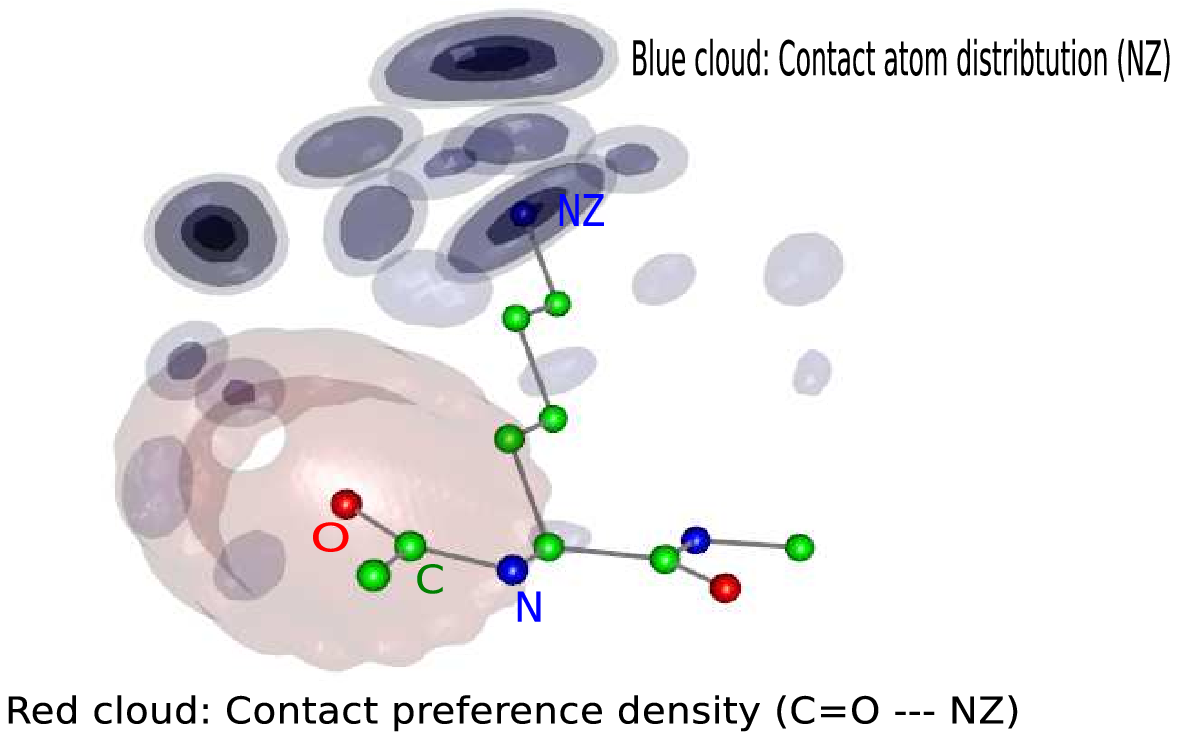}
\caption{The model system with the 3D spatial probability densities that are factors in the overlap 
density. The contact preference of the fragment (O-C-N) is depicted as a red cloud, representing a 
cutoff of the density to show the approximate form. Target, or contact, atom (NZ) distribution modelled 
as a kernel density estimate, is shown as the blue patch-like clouds with darker areas corresponding to 
higher density values. Moving between patches is realized by the side chain conformation going thermally 
around maxima like the ones in Figure \ref{fig6}.}
\label{fig8}
\end{figure}

We will now focus on the lysine side chain {NZ\textsubscript{i}} to main chain {O\textsubscript{i-1,i}} 
contact. Subscript $i$ refers to the lysine residue in the dipeptide and {O\textsubscript{i-1}} is main 
chain carbonyl oxygen of the peptide bond. Common PDB atom names are used here as identifiers. The 
dipeptide structure is depicted in Figure\ \ref{fig8} along with a modelled Target-atom 
({NZ\textsubscript{i}}) distribution and another cloud plot, for the contact preference density of 
the previous to lysine residue carbonyl. More precisely, the carbonyl group is part of the molecular 
fragment ({O\textsubscript{i-1}}-{C\textsubscript{i-1}}-{N\textsubscript{i-1}}), with respect to which 
the reference frame of the contact preference density is determined.

\subsubsection{Internal scoring} Target atom distributions are based on internaltorque 
$M(\chi_1,...,\chi_n) = M(\chi_1) + ... + M(\chi_n)$ and predefined thermal energy levels. 
Equation\ \ref{SingleT} determines a single term $M(\chi_i)$, though only component 
$M_{\alpha}(\chi)$ used, as discussed. The thermal energy cutoffs for this study were 
chosen to be 1.0 and 1.2 kcal/mol, where the latter corresponds to an average kinetic 
energy of approximatelty 4(1/2){k\textsubscript{B}}T for the side chain (four bonds 
$\chi_1,...,\chi_4$ are considered and T = 300 K). Conformations below these cutoffs might 
contain one or more less favoured rotational states about one of the four bonds, but the 
added bond-specific contributions produce the torque state, that is below the cutoff and the 
conformation is chosen for further use. In a more realistic case, these levels would be generated 
from a thermodynamic distribution, in which case the mean levels would be set according to 
temperature. Also, different degrees of freedom (rotatable bonds) can be given an individual 
response, in the form of a constraint restricting the freedom of motion, to the selected thermal 
energy. The response can for example be coupled to moment of inertia, as was done in ref 
\cite{Hakulinen2013} for an amino acid side chain. The average behaviour used in this work is 
considered to be precise enough for the purpose of presenting how the method is applied.\\

The dihedral pair ($\chi_3$,$\chi_4$)-dependent part of the side chain torque is exemplified in 
Figure\ \ref{fig6}. This torque map was calculated using M$_{\alpha}$ in Equation\ \ref{SingleT} and 
the tentative parameters given in Table \ref{BTNparams}. In the following examples, the term 
\textit{overlap} is reserved for a purely attractive interaction, and \textit{contact preference} is 
then obtained when the contact is evaluated with repulsion included.

\subsubsection{Overlap for the terminal amino nitrogen} 
\begin{figure}[ht]
\includegraphics{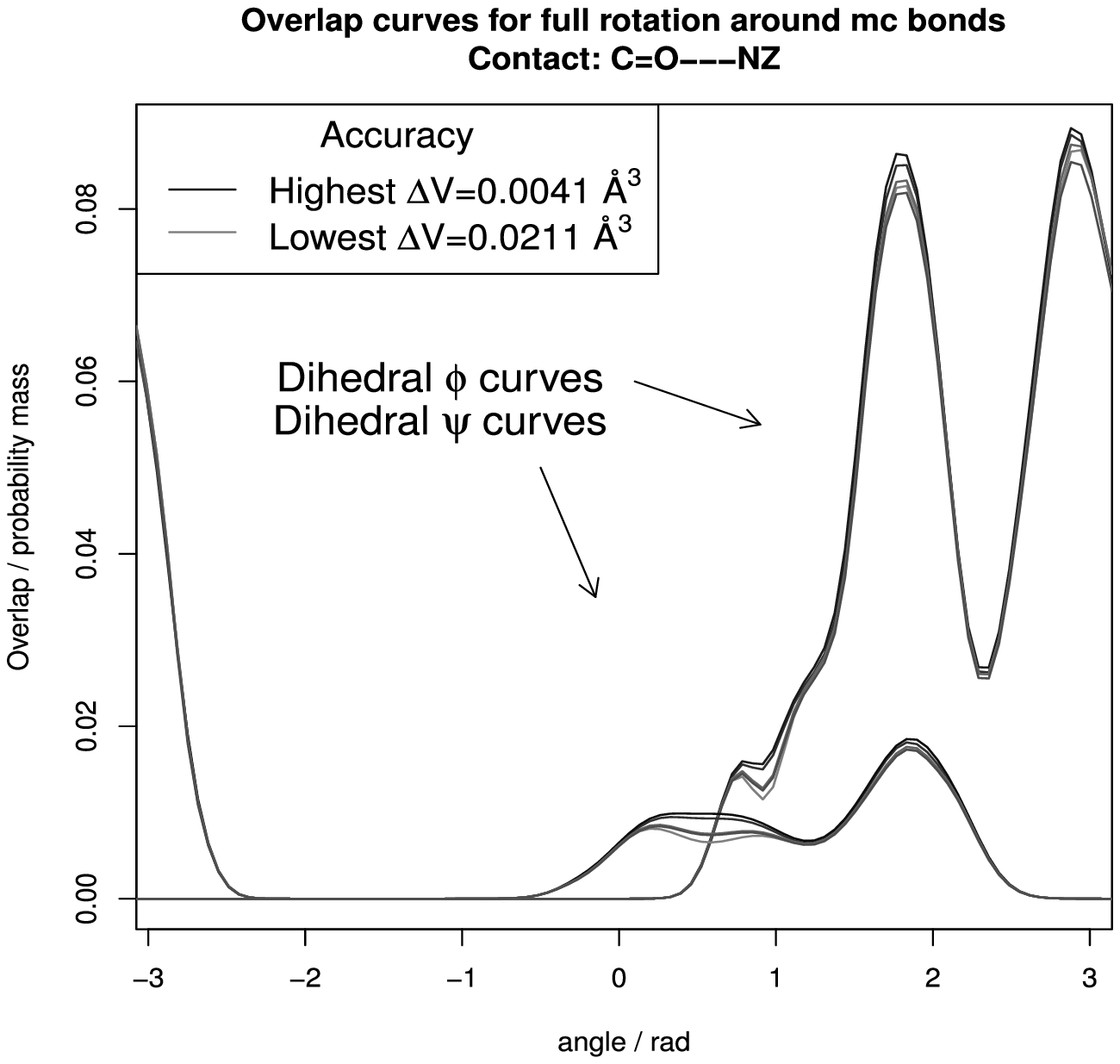}
\caption{Five calculated overlap curves over trans$\rightarrow$trans
rotation for both main chain dihedrals of the lysine, $\phi$ and $\psi$.
Standard errors along mean curves (not shown) were SE$<$0.0019 for $\phi$ and
SE$<$0.0014 for $\psi$. The inclusion of probabilistic repulsion to these
curves produces what is here called contact preference. When defining
calculation precision (pr) as the volume element size $\Delta V$, then
0.004$<$pr$<$0.021 $\AA^3$ for these curves.}
\label{fig9}
\end{figure}

The generated distribution of lysine side chain nitrogen (NZ) positions was modeled with 
a 3D kernel density estimate, Equation\ \ref{targDensity}. Values for the parameters used in 
the (main chain carbonyl oxygen) contact preference probability density, Equation\ \ref{prefDensity}, 
were estimated in our previous study \cite{Hakulinen2012}. The overlap integral was defined according 
to Equation\ \ref{OL} as an integral over the square root of the product of a kernel estimate and the 
contact preference density. Overlap values were obtained numerically as Riemann sums. The determined 
overlap profiles over a full rotation of both main chain rotatable angles $\psi$ and $\phi$ are 
shown in Figure\ \ref{fig9}.

\paragraph{Overlap profiles} The numerical overlaps over a full rotation,
as shown in Figure\ \ref{fig9}, suggest that when the main chain dihedral angle $\psi$ (C-terminus 
side) has a value in the interval [0,2.2] rad or about [0,126] degrees, the lysine side chain nitrogen 
NZ forms the strongest contact to the main chain carbonyl oxygen (O$_i$). The $\psi$ values 
around {105\textsuperscript{o}} or 1.83 rad (maximum overlap) turn the carbonyl group of the lysine 
residue to the same side with the side chain, which is a plausible conformation in order for a contact 
to form.\\

In the case of $\phi$ (N-terminus side), there are two almost equally high peaks, one at the 
same angle value as $\psi_{max}$ (1.83 rad) and the other, slightly higher peak, at 2.9 rad or 165 
degrees. The steep gradient on both sides of both the $\phi$ angle maxima, follow from directional 
preferences in the side chain to main chain contact (NZ--O) and emphasize not to rely on only a 
single main chain conformation in computer models.

\subsubsection{Side chain NZ preference for main chain C=O over full rotations}
Repulsion was defined as negative probability in the section Interaction model. It should be 
mentioned, that extreme repulsion is not possible in our model, at least not without a unrealistically 
overcrowded spatial area, since the negative probability densities that we use do not tend towards 
infinity even when two atom centers coincide. This singularity is here prevented by simultaneously 
considering the effect of the whole space occupied by the entire conformer ensemble resulting from 
thermal motion, while still assigning relatively large repulsion for close contacts.\\
\begin{figure}[ht]
\includegraphics{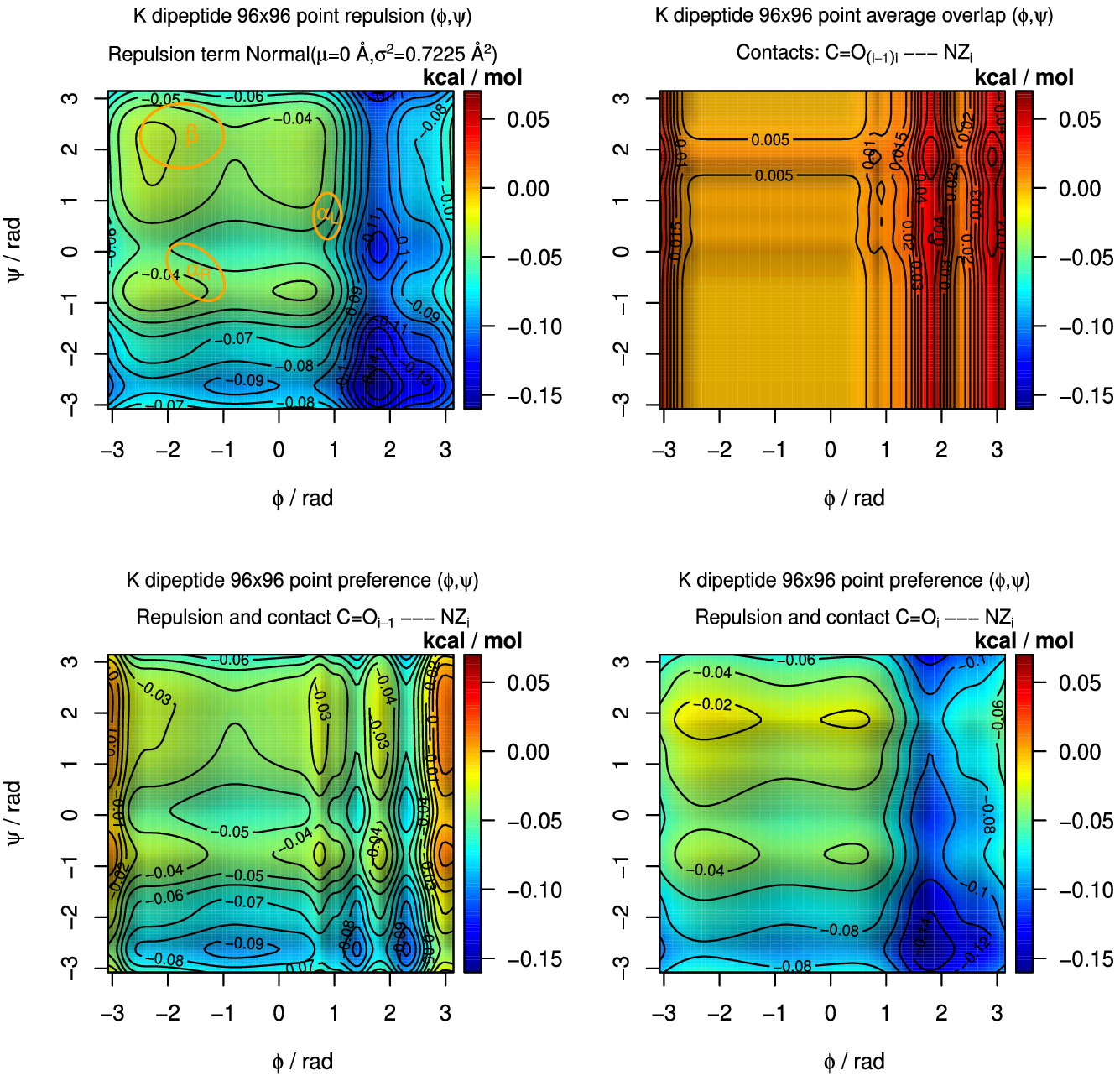}
\caption{Factors of lysine dipeptide conformational preferences. Higher value on a surface means a 
more preferred point in dihedral ($\phi$,$\psi$) space. The surfaces are repulsion map (upper left), 
average overlap (upper right), preference with contact to previous residue carbonyl (lower left) and 
preference with contact to carbonyl of the same residue (lower right). Preference is calculated by 
adding the positive overlap probability mass to the negative probability surface of repulsion. Typical 
core areas of secondary structures ($\beta$,$\alpha_R$,$\alpha_L$) are marked with labeled ellipses. On 
the lower row, the effect of overlap mass is exaggerated due to not using a contact specific weight, for 
simplicity, but the form applies.}
\label{fig10}
\end{figure}
Figure \ref{fig10} has four subfigures, and left on the upper row we show the normally distributed 
repulsion based ($\phi$,$\psi$)-map for our model system, lysine dipeptide. This map is a form of 
steric interaction based Ramachandran plot for lysine, though it should be remembered that the 
probabilistic repulsion model is a first approximation - an averaged distance dependent contact of 
main chain atoms with all generated side chain conformations simultaneously. Namely, 
($\phi$,$\psi$)-torque, directional preferences and atom type specific features are to be 
included, but already this approximation gives reasonably the basic features like typical 
beta-conformations (marked with an ellipse) in the allowed, least repulsion area, and high 
repulsion on the $\phi>$60 degrees area. The latter follows from side chain to main chain 
interactions (not observed for glycine).\\

Average overlap probability masses (0.5*OL$_{\phi}$+0.5*OL$_{\psi}$), produced by the side chain 
to main chain contacts ({NZ\textsubscript{i}} to main chain {O\textsubscript{i-1,i}}) are also 
plotted on the upper row in Figure\ \ref{fig10}. Starting from the repulsion surface on the left 
and the overlap probabilities, contact preferences (i.e., repulsion + overlap) relating to either of 
the calculated side chain to main chain interactions are shown on the lower row. Especially the 
contact that is dependent on the main chain dihedral $\phi$ ({NZ\textsubscript{i}} to 
{O\textsubscript{i-1}}, lower left), modifies strongly the repulsion based picture, which comes 
largely from that the sc to mc contacts are modelled separately as the only possible, giving them 
here the \textit{a priori} weights of one. The weigth of the overlap for this contact was in our 
earlier work about 1/7.   

\paragraph{Probability as a measure of preference}
In general, like for total energy, the overall probability mass for an interaction complex comes 
from all plausible contacts, which are considered either simultaneous or alternative. The first of 
these is calculated by adding individual overlap masses, which quantifies how preferred the 
molecular environment, like a binding pocket is. In the second case, individual overlaps are compared 
to determine strengths of the contacts, and therefore the complex specific directional preferences. 
An example is the lysine side chain nitrogen being more preferred for the $\phi$-dependent carbonyl 
oxygen than for the $\psi$-dependent. Still added, that the highest Target-atom densities obtained for 
lysine NZ are when the side chain is pointing away from the main chain part, as depicted in 
Figure\ \ref{fig8} by the blue clouds or patches. This means that contacts to the respective direction 
would get higher overlap probabilities, likely even when the contacts there would have lower 
weights C$_i$. As mentioned in the caption of Figure\ \ref{fig10}, the weights are not used in this 
work (see also Equation\ \ref{OLgen}) for simplicity, because the spatial form of the overlap 
probability density is studied in this work and it is essential to all fragment classes.\\

The interaction complex specific directional aspect of the preferences is exemplified by that, 
when overlap masses to the previous residue carbonyl are added to the repulsion map 
({NZ\textsubscript{i}} to {O\textsubscript{i-1}}, lower left in Figure\ \ref{fig10}), \emph{trans} 
conformation is strongly preferred. This is isolated dipeptide specific and not seen in Ramachandran 
plots. The same features of the lower row surfaces would be there with an \textit{a priori} weight,
but less emphasized.

\subsubsection{Referencing a quantum chemical result}
A comparison of our overlap results with those of a quantum chemical calculation was done using the 
article \cite{Zhu2012}, from which the model structure, or toy model, lysine dipeptide was picked. There 
the intrinsic energy of the dipeptide, mapped as 2D-function of first two side chain dihedrals 
($\chi_1$,$\chi_2$) showed a deep well corresponding to $\beta$ and $\alpha_L$ main chain conformations. 
These deep wells were interpreted in ref \cite{Zhu2012} as due to a hydrogen bond between side chain and 
main chain, representing a way that mc conformation affects side chain internal preferences. Both, 
$\beta$ and $\alpha_L$, were in ref \cite{Zhu2012} represented by one mc dihedral angle point, and 
for $\beta$ had been chosen the angle values ($\phi$,$\psi$)=(-120,120) degs.\\In our results, this 
mc conformation is in the band of highest $\psi$-dependent overlap probability 
mass, between 0.005-contours shown on the right on the upper row in Figure\ \ref{fig10}. The point is 
also in the most preferred area of least repulsion in ($\phi$,$\psi$)-space, when the contact from 
side chain to main chain of the same residue is added to repulsion 
({NZ\textsubscript{i}} --- {O\textsubscript{i}}, on the right on the lower row in Figure\ \ref{fig10}).\\

The other mc conformation, $\alpha_L$-helix, had in ref \cite{Zhu2012} been given the values 
($\phi$,$\psi$)=(63.5,34.8) degs, which is a location (inside $\alpha_L$-ellipse in upper left map of 
Figure\ \ref{fig10}). In the lower left map, $\phi$-dependent contact {NZ\textsubscript{i}} --- {O\textsubscript{i-1}} overlap probability mass is added to negative probability mass of the upper left map and has 
compensated some of the repulsion in the $\alpha_L$ position, from -0.07 in upper left to -0.05 in lower 
left. This is in line with the lower intrinsic energy minimum for $\alpha_L$, given in ref \cite{Zhu2012} 
(supplementary), due to a side chain main H-bond.\\The comparison is not straightforward, because in 
ref \cite{Zhu2012} had been varied two side chain dihedrals to produce the side chain conformations, 
whereas here we varied four. In addition, secondary structures were in ref \cite{Zhu2012} been 
represented by a single mc conformation and in this work full rotations were studied. Nevertheless, the 
quantum chemical work in ref \cite{Zhu2012} could be used as reference for our classical approach, which 
was one of the intended purposes of ref \cite{Zhu2012}, as stated there.\\

The surfaces in Figure\ \ref{fig10} would be altered by adding contacts to water molecules, or 
other small molecules or ions. Interactions involving these, like a water brigde, are in our 
approach modelled as mediated contacts, for an example, see \cite{Hakulinen2013}.

\subsubsection{About information content of overlap}
As presented in this work, our conformational preference model gives the internal state of a 
molecular structure as torque. The thermal environment is in this treatment incorporated into 
the predictions by forcing a thermal energy level that gives the highest accessible torque value. To 
illustrate the effect that changing a level has, full $\phi$ and $\psi$ rotation overlap sets were 
calculated with a 1.0 kcal/mol cutoff, to be compared with the 1.2 kcal/mol used for the results 
otherwise in this work.\\

\begin{figure}[ht]
\includegraphics{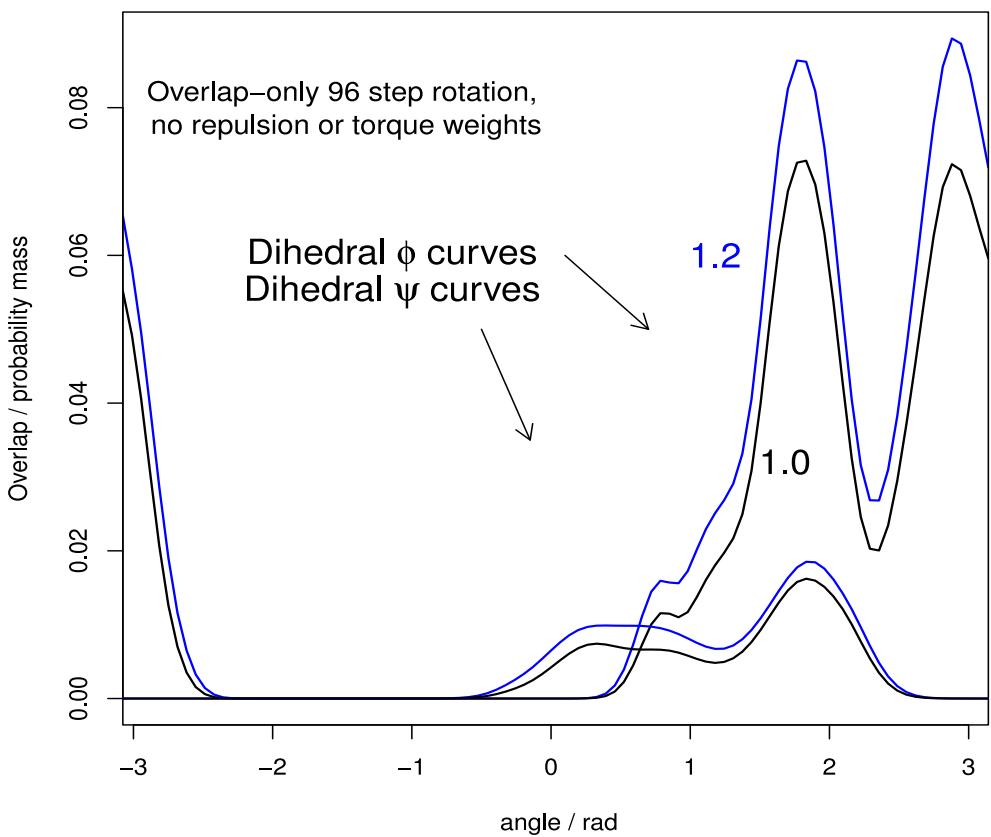}
\caption{Compared overlap values for the lysine side chain nitrogen contact with main chain carbonyl 
oxygen. Contact over the $\phi$ and $\psi$ intervals corresponding to trans $\rightarrow$ trans 
rotation. Step size $\Delta \psi$ during rotation was $\pi$/48. Two thermal energy levels, 1.0 and 1.2 
kcal/mol, used as cutoff for torque to vary accessible conformations in the target atom (NZ) distribution.}
\label{fig11}
\end{figure}

The results in Figure\ \ref{fig11} (black and blue curves) match the presumption that the lower energy 
level produces a target distribution with lower overlap than the higher level. This follows from that 
the higher energy level cutoff allows the side chain to explore more conformations, which produces a 
larger overlap with the main chain fragment contact preference density for all $\phi$ and $\psi$ values. 
Only minor deviations are seen in the relative angle dependence between these curves based on two cutoffs, 
so mainly the magnitude of overlap varies in this case. Significant changes in the overlap shape are 
possible, and would depend on the internal torque of the structure, in which case those are part of 
the correct description of the system.
\paragraph{Space and time average}The rationale behind that an overlap integral incorporates 
the effect of structural flexibility, is that because conformations are generated based on internal 
preferences, no further than 1-4 interactions, the distribution contains all (classically) accessible 
structural conformations. Further, it is taken that when this distribution is modeled as a 3D probability 
density, the functional form describes both space and time average of conformational variation. The 
overlap integrand determines then how much of the conformational space, of the external contact free 
structure, is conserved in the contact. The roles of the interacting structures can be reversed and 
must produce the same results.

\section{Discussion}
We have previously constructed a probabilistic framework for modelling molecular 
interactions \cite{Hakulinen2012}. Relating to that, we introduce in the present work a
torque-based model for scoring internal states or conformations, a negative probability 
density model for repulsive interactions, a new choice of densities to model Target-atom 
position distributions (in order to calculate overlap integrals that quantify the interactions) 
and an interpretation of the enthalpy--entropy compensation based on constant overlap probability 
mass.\\

We provide exemplifying calculations using butane, methylated butanes, alkanes like 
neopentane and a lysine dipeptide model system, in order to demonstrate the use of our method. 
The reasonableness of the numerical results obtained, using tentative parameters, was verified 
agains quantum chemical results from literature \cite{Martin2013,Zhu2012} and well-known structural 
feature like allowed areas in a Ramachandran plot and approximate energy costs for eclipsed 
conformations in hydrocarbons.\\

Numerical integration of the overlap integrals at several different levels of precision produced 
values concentrated around a specific mean value, indicating that the results are robust in the 
tested numerical precision interval. The calculated overlap profiles change sharply, due to 
directional interaction preferences, in some of the studied angle intervals, indicating therefore 
that it is insufficient to present a peptide main chain secondary structure with only one 
($\phi$,$\psi$)-conformation.\\

Further development of our method requires advances in the utilization of chemical information 
relevant for our model. A future challenge is a chemically precise molecular fragment classification. 
Such a classification may be based on quantum chemical or semi-empirical estimates for the local 
electronic structure. In the present model, contacts are classified as fragment--target interactions, 
however, a more elegant solution would be to formulate these directly as symmetric fragment--fragment 
interactions. Data collection for the improved interaction model update would then be based on the 
refined fragment classification, and the calculated electronic structure would be used to define 
new parameter charges for the torque model. In addition, combining different levels of internal motion in 
a macromolecule, in this framework, requires a separate scheme, e.g., due to different characteristic 
times of the respective conformation changes and interactions between more distant parts of the stucture.\\

In conclusion, our approach reduces the multidimensional problem of taking into account all plausible 
conformations in molecular interactions into three dimensions (at least locally) by formulating the 
interaction as an overlap integral of square roots of 3D probability densities. Therefore, this 
scoring function scheme provides an efficient way to incorporate thermal motion into a molecular model.

\section*{Acknowledgments}
This work was supported by grants from the Sigrid J\'uselius Foundation,
Tor, Pentti och Joe Borgs stiftelse at {\AA}bo Akademi University and
Medicinska Underst{\"o}dsf{\"o}reningen Liv och H{\"a}lsa rf.

\bibliographystyle{tMPH}
\bibliography{framereferences}

\label{lastpage}

\end{document}